\documentclass[fleqn,usenatbib,useAMS]{mnras}

\usepackage{graphicx} 
\usepackage{amsmath} 
\usepackage{amssymb} 
\usepackage{multicol} 
\usepackage{bm} 
\usepackage{pdflscape} 

\usepackage[T1]{fontenc}
\usepackage{ae,aecompl,lipsum}

\newcommand{\lcdm}{$\Lambda$CDM}
\newcommand{\hi}{\ion{H}{i}}
\newcommand{\om}{\Omega_{{\rm m},0}}
\newcommand{\ob}{\Omega_{{\rm b},0}}

\newcommand{\ho}{H_0}
\newcommand{\fnl}{f_{\rm NL}}
\newcommand{\ud}{{\rm d}}

\graphicspath{{./}{figures/}}

\title[Beyond $\mathit{\Lambda}$CDM with H\,{\small I} IM]{Beyond $\bf\Lambda$CDM with H\,{\Large \textbf{I}} intensity mapping: robustness of cosmological constraints in the presence of astrophysics}
\author[S. Camera \& H.\ Padmanabhan]{
Stefano Camera$^{1,2}$\thanks{stefano.camera@unito.it} and Hamsa Padmanabhan$^3$\thanks{hamsa@cita.utoronto.ca}
\\
$^1$Dipartimento di Fisica, Universit\`a degli Studi di Torino, via P.\ Giuria 1, 10125 Torino, Italy\\
$^2$INFN -- Istituto Nazionale di Fisica Nucleare, Sezione di Torino, via P.\ Giuria 1, 10125 Torino, Italy\\
$^3$Canadian Institute for Theoretical Astrophysics, 60 St.\ George Street, Toronto, ON M5S 3H8, Canada
}

\date{}

\pubyear{2019}

\begin{document}
\label{firstpage}
\pagerange{\pageref{firstpage}--\pageref{lastpage}}
\maketitle

\begin{abstract}
Mapping the unresolved intensity of the 21-cm emission of neutral hydrogen (\hi) is now regarded as one the most promising tools for cosmological investigation in the coming decades. Here, we investigate, for the first time, extensions of the standard cosmological model, such as modified gravity and primordial non-Gaussianity, taking self-consistently into account the present constraints on the astrophysics of \hi\ clustering in the treatment of the brightness temperature fluctuations. To understand the boundaries within which results thus obtained can be considered reliable, we examine the robustness of cosmological parameter estimation performed via studies of 21-cm intensity mapping, against our knowledge of the astrophysical processes leading to \hi\ clustering. Modelling of astrophysical effects affects cosmological observables through the relation linking the overall \hi\ mass in a bound object, to the mass of the underlying dark matter halo that hosts it. We quantify the biases in estimates of standard cosmological parameters and those describing modified gravity and primordial non-Gaussianity, that are obtained if one misconceives the slope of the relation between \hi\ mass and halo mass, or the lower virial velocity cut-off for a dark matter halo to be able to host \hi. Remarkably, we find that astrophysical uncertainties will not affect searches for primordial non-Gaussianity---one of the strongest science cases for \hi\ intensity mapping---despite the signal being deeply linked to the \hi\ bias.
\end{abstract}


\section{Introduction} \label{sec:intro}
Over the past decade, the standard model of cosmology, known as \lcdm, has become fairly well established, with constraints from the latest Planck survey reaching better than percent-level in the accuracy of measurement of cosmological parameters \citep{planck2018}. However, several outstanding questions remain to be answered in the context of this model. The chief two are arguably: $i)$ the nature of the late-time accelerated expansion of the cosmos, and $ii)$ the mechanism responsible for generating the primordial perturbations that led to the formation of structures. 

The former phenomenon is often attributed to dark energy, which behaves like a fluid component having negative pressure. Most observations are consistent with dark energy being a cosmological constant, however, the breakdown of general relativity on cosmological scales (i.e.\ modifications of gravity) may also explain the observed cosmic acceleration \citep[for a review, see][]{Clifton:2011jh}. Modified gravity theories thus acquire a strong significance in understanding the nature of dark energy.

On the other hand, the generation of primordial perturbations is believed to be connected to an early inflationary epoch of the Universe. The detection of primordial non-Gaussianity, easily characterizable by a nonzero value of the parameter $\fnl$ \citep[e.g.][]{Amendola:2012ys,Alonso:2015uua, Camera:2015fsa,ballardini2019mt}, places stringent constraints on theories of inflation, since standard single-field slow-roll inflationary models predict negligible non-Gaussianity \citep{Maldacena:2002vr}, while non-standard scenarios \citep[e.g.][]{Komatsu:2010fb, Bartolo:2004if}, including multi-field models, allow larger amounts. Hitherto, searches for primordial non-Gaussianity have been through measurements of higher-order correlation functions of the cosmic microwave background anisotropies \citep[e.g.][]{planck2019}, or looking for the induced scale-dependence on the bias of structures in the clustering of galaxies \citep[e.g.][]{2013MNRAS.434..684M,2014PhRvD..89b3511G, castorina2019}, whilst the multi-tracer technique is a promising new method for future surveys \citep[e.g.][]{seljak2009}.

In the past decade, a tremendous amount of theoretical interest has emerged in the field of line-intensity mapping \citep[e.g.][and references therein]{loeb2008, kovetz2019}, a novel probe of cosmology, large-scale structure, and galaxy evolution. Unlike more traditional surveys, line-intensity mapping captures the aggregate emission from a region, rather than targeting individual objects like galaxies, thus making it a fast and inexpensive approach to probe large-scale structure and evolution. The most well-studied example of line-intensity mapping is that using the redshifted 21-cm line of neutral hydrogen \citep[\hi;][]{chang10, masui13, switzer13, anderson2018}. Being a powerful probe of the largest scales of cosmological evolution, intensity mapping is thus particularly well-suited to constrain extensions to the standard \lcdm\ framework \citep[e.g.][]{Bull:2015stt}. Several studies have investigated the possibility of measuring cosmological effects at ultra-large scales using \hi\ intensity mapping \citep{Camera:2015fsa,Alonso:2015uua,Fonseca:2015laa}, and there are excellent prospects for placing stringent constraints on non-Gaussianity from current and future generation line intensity mapping surveys \citep[e.g.][]{Camera:2013kpa,racanelli2015,Alonso:2015sfa,Fonseca:2016xvi,skacosmo2018,cosmicvisions2018, bernal2019, dizgah2019}. Similarly, it is possible to place constraints on modified gravity models and the growth of structure via \hi\ intensity mapping \citep[e.g.][]{masui2010,Hall:2012wd, 2015ApJ...811..116B,pourtsidou2016, fonseca2019, castorina2019a}. 

For extracting precise and accurate cosmological information from intensity mapping surveys, an understanding of the uncertainties in the baryonic gas physics is essential. This may be referred to as the `astrophysical systematic' effect in such surveys, and plays an analogous role to other systematic effects \citep{padmanabhan2019}. A natural framework to quantify these uncertainties is the so-called `halo model approach', in which a representative set of parameters is used to capture the dominant astrophysical effects taking place in the baryonic gas. Such an approach, constrained to present-day observations of \hi\ gas over $z \sim 0-6$ \citep{Padmanabhan2017}, yields a set of 5 parameters to describe \hi\ abundances and clustering, with best-fit values and uncertainties statistically favoured by the \hi\ observations available today.

In \citet{padmanabhan2019}, the \hi\ halo model framework was used to quantify the extent by which the \textit{precision} of recovery of the standard cosmological parameters is affected by astrophysical uncertainties, assuming a standard $\Lambda$CDM cosmology. It was found that the astrophysical parameters indeed broaden the cosmological constraints, however this broadening is, in large part, mitigated by the tomographic addition of information in multiple redshift bins. 

In the present work, we extend and complement the above study, by asking the related question: how much is the \textit{accuracy} in recovery of cosmological parameters impacted by the uncertainties in \hi\ astrophysics? In other words, we calculate the relative biases of cosmological parameters, in the presence of the astrophysics of \hi\ clustering in a self-consistent manner. Furthermore, we incorporate deviations from standard $\Lambda$CDM cosmology, namely the existence of a non-Gaussianity parameter, $\fnl$, and explore the effects of modified gravity by allowing the structure growth parameter $\gamma$ to deviate from its fiducial value predicted in general relativity.

We use the cases of $\fnl$ and $\gamma$ to explore the specific impact of the \hi\ astrophysics on parameters probing beyond-\lcdm\ physics. We find that measurements of primordial non-Gaussianity are negligibly affected by the details of the astrophysical uncertainties. The framework also enables us to quantify the relative biases on standard cosmological parameters and modified gravity, that result from neglecting astrophysical uncertainties.

The paper is structured as follows. In \autoref{sec:halomodel}, we provide a brief recap of the halo model framework used for incorporating the knowledge of \hi\ astrophysics into intensity mapping surveys, and a description of the chief parameters involved in this framework. In \autoref{sec:cosmology}, we provide a brief overview of the extensions to the \lcdm\ framework, namely modifications of gravity (\autoref{sec:MG}) and primordial non-Gaussianity effects (\autoref{sec:PNG}), and illustrate how they affect the three-dimensional power spectrum of \hi. We incorporate these into the observed angular power spectrum of \hi\ (\autoref{sec:powerspectrum}) and compute the relative biases on the various cosmological and astrophysical parameters from future \hi\ intensity mapping surveys with the CHIME, MeerKAT, and SKA1-MID-like experimental configurations. We summarise our results (\autoref{sec:results}), commenting on our findings and discussing future prospects in a brief concluding section (\autoref{sec:conclusions}).

\section{Halo model of H\,{\scriptsize i} clustering}
\label{sec:halomodel}
In this section, we summarise the framework adopted to characterise the abundances and clustering of \hi\ intensity fluctuations and compute the full, nonlinear \hi\ power spectrum.

We follow the halo model approach described in \citet{Padmanabhan2017}, which uses the relation $M_{\rm HI} (M,z)$, describing the average \hi\ mass associated with a halo of mass $M$ at redshift $z$, constrained to have the form
\begin{equation}
M_{\rm HI} (M,z) = \alpha f_{\rm H,c} M \left(\frac{M}{10^{11}\,M_{\odot}/h}\right)^{\beta} 
\exp\left[-\left(\frac{v_{{\rm c},0}}{v_{\rm c}(M,z)}\right)^3\right].
\end{equation}
In the above expression, the three free parameters are given by: $\alpha$, that denotes the 
average \hi\ fraction, relative to the cosmic fraction $f_{\rm H,c}$; $\beta$, the slope of 
the \hi-halo mass relation; and $v_{{\rm c},0}$, the minimum virial velocity cut-off below which a halo would preferentially not host \hi.

The small-scale structure of the \hi\ distribution is described through the density profile of the \hi\ in the halo, $\rho_{\rm HI} (M,z)$. The best-fitting form is given by an exponential, viz.
\begin{equation}
\rho_{\rm HI} (r;M,z) = \rho_0 \exp\left[-\frac{r}{r_{\rm s}(M,z)}\right],\label{rhodefexp}
\end{equation}
where $r_{\rm s}$ is the scale radius of a dark matter halo, defined as $r_{\rm s}(M,z)\equiv R_{\rm v}(M)/c_{\rm HI}(M,z)$. Here, $R_{\rm v}(M)$ denotes the virial radius of the dark matter halo of mass $M$, and $c_{\rm HI}$ denotes the concentration of the \hi, analogous to the corresponding expression for dark matter, and is defined as \citep{maccio2007}
\begin{equation}
c_{\rm HI}(M,z) = c_{{\rm HI},0} \left(\frac{M}{10^{11} M_\odot} \right)^{-0.109} \frac{4}{(1+z)^\eta}.
\end{equation}

The ($r$-wise) constant of proportionality in \autoref{rhodefexp}, which we denote by $\rho_0$, is fixed by normalising the \hi\ mass within the virial radius $R_{\rm v}$, at a given halo mass and redshift, to be equal to $M_{\rm HI}$. This can be shown to be well approximated by
\begin{equation}
\rho_0(M,z) = \frac{M_{\rm HI} (M)}{8 \pi r_{\rm s}^3(M,z)}.\label{eq:rho0}
\end{equation}
The \hi\ radial density profile thus introduces two additional free parameters: the concentration normalisation, $c_{\rm HI,0}$; and its redshift slope, $\eta$.\footnote{Note that the slope $\eta$ is often denoted by $\gamma$ in the literature. We choose to change this symbol to avoid confusion with the modified gravity parameter $\gamma$ (see \autoref{sec:MG}).}

 The above framework has been calibrated by \citet{Padmanabhan2017} using a Markov Chain Monte Carlo procedure, to the combination of all the \hi\ data available in the literature today, which includes the 21-cm intensity mapping experiments \citep[e.g.][]{switzer13}, DLA observations across $z \sim 2-5$ \citep[e.g.][]{rao06, zafar2013, fontribera2012, noterdaeme12} and \hi\ galaxy surveys around $z \sim 0-1$ \citep{zwaan05, zwaan2005a, martin12}. For a full list, see Table 2 of \protect\citealt{Padmanabhan2017}. The form of the \hi-halo mass relation inferred is also supported by the
results of simulations \citep[e.g.][]{villa14}, which
find that almost all ($\gtrsim 90\%)$ of the \hi\ in the post-reionisation ($z < 6$) Universe resides within haloes.

The best-fit values and $68\%$ confidence intervals for the five free parameters in the halo model obtained in \citet{Padmanabhan2017} by fitting to the combination of high- and low-$z$ \hi\ data take the following values
\begin{align}
\log v_{{\rm c},0} &= 1.56 \pm 0.04,\nonumber\\
\eta &= 1.45 \pm 0.04,\nonumber\\
c_{\rm HI,0} &= 28.65 \pm 1.76,\nonumber\\
\alpha &= 0.09 \pm 0.01,\nonumber\\
\beta &= -0.58 \pm 0.06.
\end{align}

To define the power spectrum of \hi\ intensity fluctuations, we need the normalised first-order Hankel transform of the \hi\ density profile, denoted by $u_{\rm HI}(k|M,z)$ and defined as
\begin{equation}
u_{\rm HI}(k|M,z) = \frac{4 \pi}{M_{\rm HI} (M)} \int_0^{R_{\rm v}(M)}\ud r\, \rho_{\rm HI}(r;M,z) \frac{\sin (kr)}{kr} r^2,
\end{equation}
where the normalisation is to the total \hi\ mass in the halo. Note that the profile is actually truncated at the virial radius of the host halo. Using \autoref{rhodefexp} and \autoref{eq:rho0}, we find
\begin{equation}
u_{\rm HI}(k|M,z) = \left[1 + k^2 r_{\rm s}^2(M,z)\right]^{-2}.
\end{equation}

The full, nonlinear \hi\ power spectrum contains both 1-halo and 2-halo terms. The latter, denoted by $P_{\rm HI, 2h}$ accounts for correlations in the clustering of \hi\ from two distinct dark matter haloes. This term therefore primarily feels the large-scale gravitational potential of the cosmic web to which the dark matter haloes belong; it also encodes information on the bias of \hi\ with respect to the underlying matter distribution. 
The 2-halo term can be expressed as
\begin{multline}
P_{\rm HI, 2h} (k,z) = P_{\rm lin}(k,z)\\
\times\left[
\int \ud M\,n_{\rm h}(M, z) b_{\rm h}(M, z, k) \frac{M_{\rm HI} (M)}{\bar{\rho}_{\rm HI}(z)} |u_{\rm HI} (k|M,z)| \right]^2,\label{eq:PHI2h}
\end{multline}
where we integrate over the halo mass, $M$, and $n_{\rm h}$ is the halo mass function; $b_{\rm h}$ is the corresponding halo bias, and $P_{\rm lin}$ is the linear matter power spectrum.

On the other hand, the 1-halo term, denoted by $P_{\rm HI, 1h}(k,z)$, describes correlations between \hi\ structures within the same dark matter halo. Hence, this term is more affected by the dynamics on smaller, mildly to highly nonlinear scales, and it is dependent on the density profile of \hi\ within dark matter haloes. This term is given by
\begin{equation}
P_{\rm HI, 1h}(k,z) = \int \ud M \, n_{\rm h}(M, z) \left[\frac{M_{\rm HI} (M)}{\bar{\rho}_{\rm HI}(z)}\right]^2 \ |u_{\rm HI} (k|M,z)|^2.\label{eq:PHI1h}
\end{equation}

In both expressions above, $\bar{\rho}_{\rm HI}(z)$ is the mean \hi\ density in the Universe at redshift $z$, given by
\begin{equation}
\bar{\rho}_{\rm HI}(z) = \int \ud M\, n_{\rm h}(M,z) M_{\rm HI} (M).
\end{equation}
Similarly, we can define an effective \hi\ bias on large scales according to
\begin{equation}
b_{\rm HI}(z) = \int \ud M\, b_{\rm h}(M,z)n_{\rm h}(M,z) \frac{M_{\rm HI} (M)}{\bar{\rho}_{\rm HI}(z)},
\end{equation}
which is basically the content of the square brackets in \autoref{eq:PHI2h} in the $k\to0$ limit---a regime in which the (aptly normalised) $u_{\rm HI} (k|M,z)$ tends to unity and can therefore filter out of the integral. It is easy to see that this reconciles the halo model formalism to the usual approximation $P_{\rm HI}=b_{\rm HI}^2P_{\rm lin}$, valid on large scales.

Finally, the full power spectrum of \hi\ fluctuations is the sum of the one- and two-halo terms:
\begin{equation}
P_{\rm HI}(k,z) = P_{\rm HI, 1 h}(k,z) + P_{\rm HI, 2h}(k,z) 
\end{equation}
and thus contains both cosmological as well as astrophysical information from \hi\ in galaxies.

\section{Cosmological information}
\label{sec:cosmology}
In the halo model approach described in \autoref{sec:halomodel}, the information on the cosmological model is encoded in three quantities: the linear matter power spectrum, $P_{\rm lin}$; the halo mass function, $n_{\rm h}$; and the halo bias, $b_{\rm h}$. Here, we use the publicly available code \texttt{CAMB} \citep{Lewis:1999bs} to compute $P_{\rm lin}$ for a given cosmology, whilst for $n_{\rm h}$ and $b_{\rm h}$ we follow \citet{Sheth:2001dp} and \citet{scoccimarro2001}, respectively. 

Below, we briefly review the cosmological parameters involved in the analysis, for the \lcdm\ model, as well as for some of its most compelling extensions currently under investigation by the community: modified gravity and primordial non-Gaussianity \citep[for a review, see][]{Bull:2015stt}.

\subsection{$\bf\Lambda$CDM cosmology}
The standard cosmological model in our analysis is represented by a flat, `vanilla' \lcdm\ Universe with parameters: $\om=0.281$, denoting today's total matter fraction; $\ob=0.0462$, the fraction of baryons at the present time; $h\equiv\ho/(100\,\mathrm{km\,s^{-1}\,Mpc^{-1}})=0.71$, the dimensionless Hubble constant; $n_{\rm s}=0.963$, the slope of the primordial power spectrum; $\sigma_8=0.8$, the normalisation of the present-day matter power spectrum in terms of its rms fluctuations in spheres of $8\,h^{-1}\,\mathrm{Mpc}$ radius.

\subsection{Modified gravity}\label{sec:MG}
A possible solution to the long-standing `cosmological constant problem', related to the nature of the late-time accelerated cosmic expansion, is represented by modified gravity. In other words, the effects we ascribe to the mysterious component dubbed dark energy may in fact be due to a departure, on cosmological scales, of the behaviour of gravity from Einstein's general relativity (see \citealt{Clifton:2011jh} for a thorough review on modified gravity, and \citealt{Amendola:2012ys,Amendola:2016saw} for an exhaustive list of such models of interest for cosmology).

One of the main features of modified gravity is that it can alter the growth of cosmic structures, even being indistinguishable from \lcdm\ at background level. Although each modified gravity theory predicts a different growth of structure, a convenient way of parameterising such deviations from general relativity is through a generalisation of the growth rate, $f(z)\equiv-\ud\ln D(z)/\ud\ln(1+z)$, where $D(z)$ is the growth factor (normalised such that $D\to1$ for $z\to0$). In general relativity, a very good approximation is known to be $f(z)=[\Omega_{\rm m}(z)]^\gamma$, with $\Omega_{\rm m}(z)=\om(1+z)^3\ho^2/H^2(z)$ and $\gamma\approx0.55$ \citep{Lahav:1991wc}. If we now promote $\gamma$ to be a free parameter, any measured value different from $0.55$ at a statistically significant level will be a hint of modified gravity \citep{Linder:2005in}.

It is easy to see how such a modification of gravity impacts the clustering of \hi. The response of the \hi\ power spectrum to the modified gravity parameter $\gamma$ can be defined as
\begin{equation}
\frac{\partial P_{\rm HI}(k,z)}{\partial\gamma} = \frac{\partial P_{\rm HI, 1h}(k,z)}{\partial\gamma}+\frac{\partial P_{\rm HI, 2h}(k,z)}{\partial\gamma}.
\label{eq:dphidgamma}
\end{equation}
From the definition of $f(z)$ we then have
\begin{multline}
\frac{\partial P_{\rm HI, 1h}(k,z)}{\partial\gamma} = -\frac{\partial\ln D}{\partial\gamma}(z)\\
\times\int \ud M \, \frac{\partial\ln n_{\rm h}}{\partial\ln\nu}(M,z)n_{\rm h}(M,z)\left[\frac{M_{\rm HI} (M)}{\bar{\rho}_{\rm HI}(z)}\right]^2 \ |u_{\rm HI} (k|M)|^2,
\end{multline}
and
\begin{multline}
\frac{\partial P_{\rm HI, 2h}(k,z)}{\partial\gamma} = 2\frac{\partial\ln D}{\partial\gamma}(z)\bigg\{P_{\rm HI, 2h}(k,z)\\
-P_{\rm lin} (k,z)\left[
\int \ud M \, n_{\rm h}(M, z) b_{\rm h}(M, z) \frac{M_{\rm HI} (M)}{\bar{\rho}_{\rm HI}(z)} |u_{\rm HI} (k|M)| \right]\\
\times\int \ud M \,\left[\frac{\partial\ln n_{\rm h}}{\partial\ln\nu}(M,z)+\frac{\partial\ln b_{\rm h}}{\partial\ln\nu}(M,z)\right]\\
\times n_{\rm h}(M, z) b_{\rm h}(M, z) \frac{M_{\rm HI} (M)}{\bar{\rho}_{\rm HI}(z)} |u_{\rm HI} (k|M)|
\bigg\}.
\end{multline}
Here, we have used the chain rule to write the $\gamma$-dependence of both the halo mass function and the halo bias in terms of their dependence upon the over-density threshold $\nu(M,z)=\delta_{\rm c}/D(z)/\sigma(M)$, with $\delta_{\rm c}\simeq1.69$ the critical density for spherical collapse, and $\sigma^2(M)$ the mass variance.

\subsection{Primordial non-Gaussianity}\label{sec:PNG}
Most inflationary scenarios predict that the distribution of density perturbations at the end of inflation contained a small degree of non-Gaussianity, usually parameterised by its amplitude, $\fnl$. Such non-Gaussianity manifests itself as a non-vanishing bispectrum of perturbations at early times; not having measured it in the cosmic microwave background has allowed us to put upper bounds on the value of $\fnl$. However, any $\fnl\ne0$ will also induce a peculiar scale-dependence on the halo bias, mostly on large scales. In particular, the local-type, scale-dependent modification to the scale-independent Gaussian linear bias of haloes reads \citep{Dalal:2007cu,Verde:2009hy}
\begin{equation}
\Delta b_{\rm h}(M,z,k)=\frac{3g_\infty\om\ho^2\delta_{\rm c}}{g_0D(z)T(k)k^2}\left[b_{\rm h}(M,z)-1\right]\fnl,\label{eq:delta_b}
\end{equation}
where $T(k)$ is the matter transfer function (normalised such that $T\to1$ for $k\to0$), and $g_0$ and $g_\infty$ are, respectively, the present-day and primordial values of the growth factor of the gravitational potential, $g(z)\propto(1+z)D(z)$. The total halo bias is then the sum of the Gaussian and non-Gaussian contributions, i.e.\ $b_{\rm h}(M,z)+\Delta b_{\rm h}(M,z,k)$.

From this it is also straightforward to derive the response of the \hi\ power spectrum with respect to $\fnl$, when the fiducial cosmology is $\fnl=0$, i.e.\
\begin{multline}
\frac{\partial P_{\rm HI} (k,z)}{\partial\fnl} = \frac{2}{\fnl}P_{\rm lin} (k,z)\\
\times\left[
\int \ud M \, n_{\rm h}(M, z) b_{\rm h}(M, z) \frac{M_{\rm HI} (M)}{\bar{\rho}_{\rm HI}(z)} |u_{\rm HI} (k|M,z)| \right]\\
\times\left[\int \ud M \, n_{\rm h}(M, z) \Delta b_{\rm h}(M, z, k) \frac{M_{\rm HI} (M)}{\bar{\rho}_{\rm HI}(z)} |u_{\rm HI} (k|M,z)| \right].
\label{eq:dphidfnl}
\end{multline}

\section{The observed H\,{\scriptsize i} angular spectrum}
\label{sec:powerspectrum}
In our analysis, we focus on the power spectrum of \hi\ fluctuations in harmonic space, $C_\ell^{\rm HI}$, which is often referred to as the angular power spectrum. To compute the angular power spectrum between two thin redshift slices centred at $z$ and $z^\prime$, we use the standard expression
\begin{multline}
C_\ell^{\rm HI}(z,z^\prime) = \frac {2} {\pi} \int \ud\tilde z\, W_{\rm HI}(\tilde z) \int \ud\tilde z^\prime\, W_{\rm HI}(\tilde z^\prime)\\
\times\int \ud k\, k^2 \left\langle \delta_{\rm HI} 
({\bm k}, \tilde z) \delta_{\rm HI} ({\bm k}^\prime, \tilde z^\prime) \right\rangle 
j_\ell\left[k\chi(\tilde z)\right] j_\ell\left[k\chi(\tilde z^\prime)\right],
\label{eq:lintheory}
\end{multline}
Here, $\langle \delta_{\rm HI} ({\bm k}, \tilde z) \delta_{\rm 
HI} ({\bm k}^\prime, \tilde z^\prime) \rangle$ is the ensemble average of the \hi\ density 
fluctuations at $({\bm k},\tilde z)$ and $({\bm k}^\prime, \tilde z^\prime)$, respectively. This quantity is, in general, not expressible purely in terms of the power spectrum of \hi\ as 
defined above, $P_{\rm HI} (k,z)$, at either of $z$ or $z^\prime$, because the underlying density field $\delta({\bm k},z)$ evolves with $z$; and so does $\delta_{\rm HI} ({\bm k}, z)$. However, if one is considering very narrow redshift bins, this can be approximated as $P_{\rm HI} (k,z_{\rm m})$, where $z_{\rm m}$ is the mean redshift of the given bin. In what follows, we use $z$ and $z_{\rm m}$ interchangeably, noting that we consider the limit of narrow bins of width $\Delta z=0.05$. Then, $W_{\rm HI}$ is the \hi\ window function in the redshift bin(s) considered, and we assume it to be uniform across the bin. Finally, $\chi(z)$ is the co-moving distance to redshift $z$, and $j_\ell$ is the spherical Bessel function of order $\ell$.

In the present study, we employ the so-called Limber approximation to simplify the integral in \autoref{eq:lintheory} for the large-$\ell$ limit, which then reads
\begin{equation}
C_\ell^{\rm HI}(z) \simeq \frac{1}{c} \int \ud z\, \frac{W_{\rm HI}(z)^2 H(z)}{\chi(z)^2} P_{\rm HI} \left[\frac{\ell}{\chi(z)}, z\right].
\label{cllimber}
\end{equation}
The Limber approximation is expected to be accurate to within 1\% above $\ell \sim 10$, for the case of narrow redshift bins such as ours (see \citealt{padmanabhan2019} and Fig.~1 of \citealt{LoVerde:2008re}) at redshifts similar to the ones we consider here.\footnote{Since primordial non-Gaussianity effects are strongest on the largest cosmic scales, the Limber approximation is customarily dropped in (either actual or forecast) measurements of $\fnl$. However, we are here interested in the \textit{relative bias} on its measurement, and the same induced systematic on $\sigma$ will reflect on $b$, too, leading to an overall qualitative cancellation in $b/\sigma$. Moreover, we stress that the Limber approximation tends to over-weigh small multipoles, which translates into a spurious enhancement of the effect of $\fnl$. For this reason, our results are conservative, as they, at most, over-estimate the constraining power on primordial non-Gaussianity.} Note that we do not include further terms in the angular power spectrum such as redshift-space distortions, for their nonlinear modelling in harmonic space is still highly uncertain \citep{Jalilvand:2019brk}, whereas their treatment in the Limber approximation on linear scales has recently been developed \citep{2019MNRAS.489.3385T}.

Noise on the angular power spectrum is calculated following \citet[][see also \citealt{Knox:1995dq, jalilvand2019, bull2014}]{ballardini2019}. In particular, for interferometers such as CHIME, we adopt
\begin{equation}
N_\ell^{\rm HI,int} = \frac{4 \pi f _{\rm sky}}{{\rm FoV}n_{\rm base}(u)n_{\rm pol} N_{\rm beam}  t_{\rm tot} \Delta \nu} \left(\frac{\lambda_{\rm obs}^2}{A_{\rm eff}}\right)^2\left(\frac{T_{\rm sys}}{\bar{T}}\right)^2.
\label{noise_int}
\end{equation}
In the above expression: $f_{\rm sky}$ is the fraction of sky covered by the survey; the field of view for interferometric dishes is approximated as ${\rm FoV} \approx \pi/2 \lambda_{\rm obs}/W_{\rm cyl}$ for this configuration \citep{newburgh2014}, with $\lambda_{\rm obs}$ the observed (i.e.\ redshifted) wavelength and $W_{\rm cyl} = 20\,\mathrm{m}$ the width of each cylinder; $n_{\rm pol}=2$ is the number of polarisations; $n_{\rm base}(u)$ is the baseline number density, expressible in terms of the multipole $\ell$ via $u=\ell/(2\pi)$; $N_{\rm beam}=N_f N_{\rm cyl}$ is the number of independent beams, where $N_f = 256$ is the number of feeds and $N_{\rm cyl} = 4$ the number of cylinders; $\Delta \nu$ is the frequency band channel width, which is connected to the tomographic redshift bin separation, $\Delta z$, by $\Delta \nu = \nu_{\rm HI} \Delta z/(1+z)^2$, and $\nu_{\rm HI}=1420\,\mathrm{MHz}$; and $A_{\rm eff} = \eta\,L_{\rm cyl}W_{\rm cyl}N_{\rm cyl}/N_{\rm feed}$ is the effective area, with $L_{\rm cyl} = 100\,\mathrm{m}$ the length of each cylinder and the aperture efficiency denoted by $\eta$ and assumed to be $0.7$ \citep{jalilvand2019}. Then, $T_{\rm sys}$ denotes the system temperature, calculated following $T_{\rm 
sys} = T_{\rm inst} \ + \ 60 \ {\rm K} \left(\nu/300 \ {\rm MHz}\right)^{-2.5}$, with $T_{\rm inst}$ is the instrumental temperature, and $\nu$ denotes 
the observing frequency, whilst $\bar T$ is the mean brightness temperature, depending on redshift $z$ according to
\begin{equation}
\bar T(z) \simeq 44\ho \left(\frac{\Omega_{\rm 
HI}}{2.45\times10^{-4}\,h^{-1}} \right)\frac{(1+z)^2}{H(z)}\,\mathrm{\mu K}.
\label{eq:tbar}
\end{equation}
For the purposes of the noise calculation, we also assume that $\Omega_{\rm HI} = 2.45\times10^{-4}\,h^{-1}$, independent of redshift. The baseline number density is approximated as independent of $u$ up to a maximum baseline length $u_{\rm max}$, given by $n_{\rm base}(u) = N_{\rm beam}^2/(2 \pi u_{\rm max}^2)$, with $u_{\rm max}$ the longest baseline $d_{\rm max}=269\,\mathrm{m}$ measured in wavelength units \citep{obuljen2018}, namely $u_{\rm max} = d_{\rm max}/\lambda_{\rm obs}$.

Instead, in the case of dish surveys like MeerKAT and SKA1-MID, we have
\begin{equation}
N_\ell^{\rm HI,dish} = \frac{W _{\rm beam}^2(\ell)}{2N_{\rm dish}  t_{\rm pix} \Delta \nu} \left(\frac{T_{\rm sys}}{\bar{T}}\right)^2 
\left(\frac{\lambda_{\rm obs}}{D_{\rm dish}}\right)^2,
\label{noise_dish}
\end{equation}
where
\begin{equation}
W _{\rm beam}^2(\ell)=\exp\left[\frac{\ell(\ell+1)\theta_{\rm beam}^2}{8\ln2}\right],
\label{W_beam}
\end{equation}
accounts for beam smoothing, with $\theta_{\rm beam}\approx\lambda_{\rm obs}/D_{\rm dish}$. Then, $N_{\rm dish}$ is the number of dishes, and each of them is assumed to have the diameter $D_{\rm dish}$, and the integration time per pixel is denoted by $t_{\rm pix}$ (taken to be 1 year for all the surveys considered here).

The standard deviation of the angular power spectrum is denoted by $\Delta C_\ell^{\rm HI}$ and defined through
\begin{equation}
\Delta C_\ell^{\rm HI} = \sqrt{\frac{2}{(2 \ell + 1)f_{\rm sky}}} \left(C_\ell^{\rm HI} + N_\ell^{\rm HI}\right),\label{eq:Delta_Cl}
\end{equation}
where  we have omitted redshift dependencies for simplicity. Note, however, that noise on the angular power spectrum effectively depends on redshift, both explicitly---as is the case of the cosmological evolution of $\bar T$---and implicitly---because of the various frequency-dependent quantities entering \autoref{noise_int} and \autoref{noise_dish}.

\subsection{Fisher matrix methodology}

\begin{figure*}
    \centering
    \includegraphics[width=\textwidth]{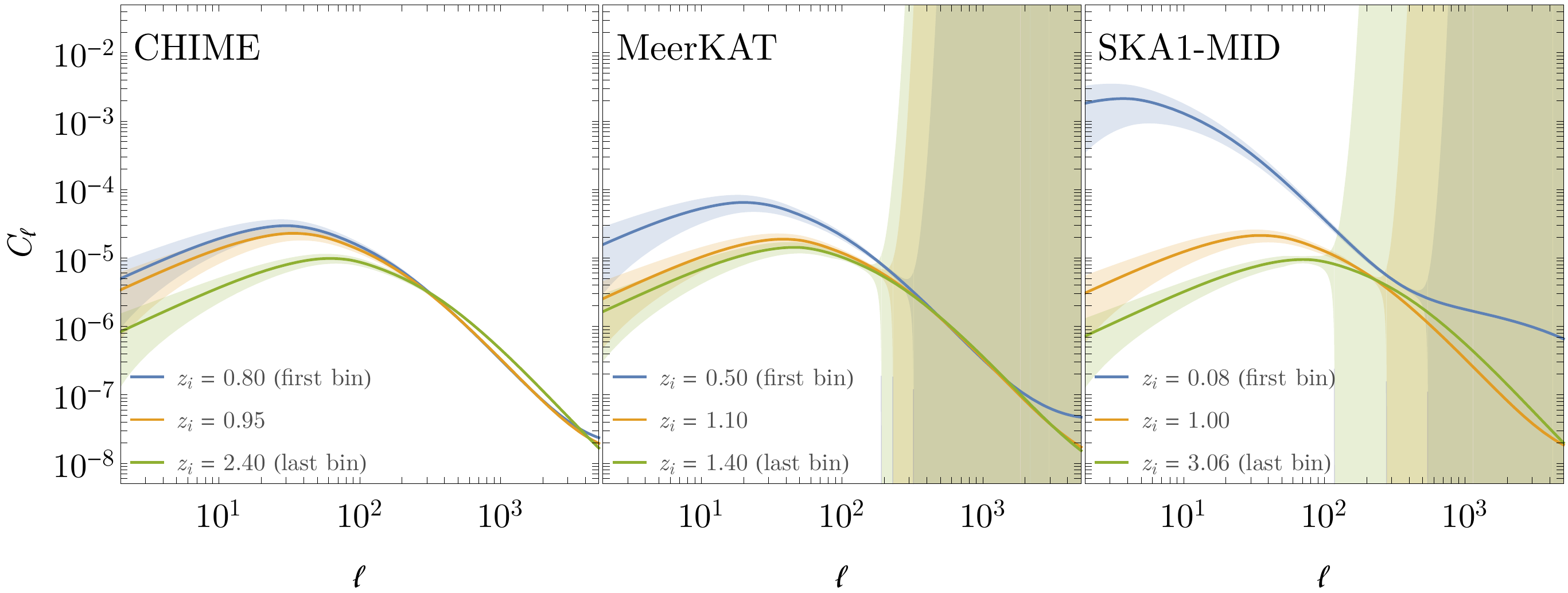}
    \caption{Fiducial angular power spectra (solid curves) and per-multipole $1\sigma$ uncertainties (shaded envelopes) for the three experimental set-ups in consideration (CHIME, MeerKAT and SKAI-MID from left to right) and three reference redshift bins, the first (blue), the last (green), and an intermediate bin around $z\sim1$ (orange).}
    \label{fig:Cl_comparison}
\end{figure*}
Having computed the angular power spectra $C_\ell^{\rm HI}$ and noise $N_\ell^{\rm HI}$, in the present section we construct Fisher matrix forecasts for the bias on cosmological and astrophysical parameters. The fiducial parameter values and redshift ranges for the experiments under consideration are provided in \autoref{table:experiments}.
\begin{table*}
\centering
\caption{Details of the experimental configurations considered in this work.}
\begin{tabular}{ccccccl}
\hline 
Configuration & $T_{\rm 
inst}\,\mathrm{[K]}$ & $N_{\rm dish} $ & $D_{\rm dish}\,\mathrm{[m]}$ & $f_{\rm sky}$ & $d_{\rm max}$ [m] & \multicolumn{1}{c}{Redshift bin centres, $z_i$}\\
\hline
CHIME & $50$ & $1024{\color{red} ^\dag}$ & $20$ & $0.61$ & 269 & $\{0.8, 0.95, 1.2, 1.35, 1.6, 1.75, 2.0, 2.25, 2.4\}$
\\ 
MeerKAT & $29$ & $64$ & $13.5$ & $0.61$ & 1000 & $\{0.5, 0.65, 0.8, 0.85, 1.1, 1.25, 1.4\}$ \\
SKA1-MID & $28$ & $190$ & $15$ & $0.61$ & 1000 &
$\{0.082, 0.35, 0.58, 1., 1.5, 2., 2.3, 3.06\}$ \\
\hline
\end{tabular}\label{table:experiments}\\
\raggedright\footnotesize{$^\dag$ For CHIME, this is the number of beams, $N_{\rm beam}=N_fN_{\rm cyl}$.}
\end{table*}

Before delving deeper into the Fisher matrix analysis, though, let us get acquainted with the power spectra involved and their measurement uncertainty. In \autoref{fig:Cl_comparison}, the three panels from left to right refer to CHIME, MeerKAT, and SKA1-MID. In each of them, the three solid curves depict the theoretical expectation in a reference bin around $z=1$ (orange colour), as well as in the first and last bin of each experimental configuration (blue and green colour, respectively). The shaded area surrounding each curve marks the $1\sigma$ uncertainty as computed via \autoref{eq:Delta_Cl}. Some key information can easily be extracted from this plot. First, as expected, the clustering signal decreases with redshift; and indeed all panels share the same blue-orange-green hierarchy, from top to bottom. Secondly, the lower the redshift, the more important the nonlinear effects; this is confirmed by comparing the small scale behaviour of the curves of different colours, with the low-redshift, blue ones hinting at the 1-halo ramp-up in power---when not overtly showing it, as in the case of SKA1-MID. Finally, the beam effects of non-interferometric experiments smooth all scales below the resolution, set by $\theta_{\rm beam}$; accordingly, the shaded regions in the central and rightmost panels start to blow up at approximately $\pi/\theta_{\rm beam}(z_1)$, with $z_1$ be the value of the central redshift of the first bin (as also indicated in the rightmost column of \autoref{table:experiments}).

We specifically focus on computing the Fisher bias on $\fnl$ and $\gamma$ due to incorrect assumptions on the astrophysical parameters. Entries of the Fisher matrix in the $i$th redshift bin and at a given multipole take the form
\begin{equation}
F_{\mu\nu,\ell} (z_i) = \frac{\partial_\mu C_\ell^{\rm HI}(z_i)
\partial_\nu C_\ell^{\rm HI}(z_i)}{\left[\Delta C_\ell^{\rm HI}(z_i)\right]^2},\label{eq:Fisher}
\end{equation}
where $\partial_\mu$ denotes the partial derivative taken with respect to parameter $p_\mu$. Of the astrophysical parameters, we consider variations only in $\beta$ and $v_{\rm c,0}$, which are found to be primarily constrained by \hi\ intensity mapping studies \citep[see][]{Padmanabhan2017}.\footnote{In practice, we adopt the decadic logarithm of $v_{\rm c,0}$ (in units of $\mathrm{km\,s^{-1}}$) as a parameter, to reduce the dynamic range of the Fisher matrices and thus ensure their numerical stability \citep[see also][]{Camera:2018jys}.} The derivatives $\partial_\mu C_\ell^{\rm HI}(z_i)$ for $p_\mu$ $\in \{h, \om, \ob, n_{\rm s}, \sigma_8, v_{\rm c,0}, \beta \}$ are computed directly from the \hi\ halo model, whilst derivatives with respect to the modified-gravity parameter $\gamma$ and the non-Gaussianity parameter $\fnl$ are analytical, given by integrating \autoref{eq:dphidgamma} and \autoref{eq:dphidfnl}, respectively, according to \autoref{cllimber}.\footnote{Note that modified gravity and primordial non-Gaussianity are considered as two different extensions of the concordance cosmological model, i.e.\ either $\gamma$ or $\fnl$ are present in the Fisher matrix at the same time.} A summary of the astrophysical and cosmological parameters and their fiducial values is provided in \autoref{table:parameters}.
\begin{table}
\centering
\caption{Fiducial values of astrophysical and cosmological parameters considered in the Fisher analysis. The astrophysical parameters are from the best-fitting \hi\ halo model \citep{Padmanabhan2017}. The cosmological parameters are in good agreement with most available observations, including the \textit{Planck} results \citep{Ade:2015xua}.}\label{table:parameters}
\begin{tabular}{clcl}
\hline
\multicolumn{2}{c}{Astrophysical} & \multicolumn{2}{c}{Cosmological}\\
\hline
$v_{\rm c,0}\,\mathrm{[km\,s^{-1}]}$ & $10^{1.56}$ & $h$ & $0.71$ \\
$\beta$ & $-0.58$ & $\om$ & $0.28$ \\
$\alpha$ & $0.09$ & $\ob$ & $0.0462$ \\
$c_{\rm HI, 0}$ & $28.65$ & $\sigma_8$ & $0.8$ \\
$\eta$ & $1.45$ & $n_{\rm s}$ & $0.963$ \\
& & $\fnl$ & $0$ \\
& & $\gamma$ & $0.55$ \\
\hline
\end{tabular}
\end{table}

For each experimental configuration, we compute the $C^{\rm HI}_\ell$, $N^{\rm HI}_\ell$, and $\Delta C_\ell^{\rm HI}$ on range of representative redshifts, $z_i$, $i=1\ldots N_z$ (see \autoref{table:experiments}) and at 15 equi-log spaced $\ell$ values between $\ell = 1$ and 1000. The $z$-bin centres are spaced by $3-5$ times the width of each individual bin in order to minimise correlations between bins (which we do not take into consideration). This is further justified by looking at Fig.~2 of \citet{Camera:2018jys}, where it is shown that the correlation between neighbouring top-hat redshift bins quickly falls as a function of the bin width, implying that most of the information comes from correlations at small radial separation. 15 such equi-log spaced multipoles are used as the edges of 14 $\ell$-bins, with the width of the $m$th bin given by $\Delta \ell_m = \ell_{m+1} - \ell_m$, $m = 1\ldots N_\ell$, defining $N_\ell=15-1=14$. For each redshift bin and pair of parameters, $F_{\mu\nu,\ell}(z_i)$ of \autoref{eq:Fisher} is linearly interpolated in $\ell$-space, and then combined to construct the full, cumulative Fisher matrix according to
\begin{equation}
\mathbfss F = \sum_{i=1}^{N_z}\sum_{m=3}^{N_\ell} \Delta\ell_m \mathbfss F_{\bar{\ell}_m}(z_i),\label{eq:F_tot}
\end{equation}
where $\bar{\ell}_m$ is the (logarithmic) centre of the $m$th multipole bin. Note that we sum over $m\ge3$ so as to include neither the monopole nor the dipole in our analysis.

From $\mathbfss F$, we can compute the forecasted marginalised errors on the various parameters as
\begin{equation}
\sigma_{p_\mu} = \sqrt{\left(\mathbfss F^{-1}\right)_{\mu\mu}},
\end{equation}
where we first take the inverse of the full Fisher matrix, and then select (the square root of) the diagonal elements. To compute the biases on the forecasted cosmological parameters induced by a wrong assumption on the astrophysical ones, we follow the treatment in \citet[][Appendix~A]{Camera:2016owj} which is based on the formalism of \citet{Heavens:2007ka} and clarifies the role of the parameters related to systematic effects. In this framework, often called as `nested likelihoods', we remind the reader that it is necessary to split the parameter set into two sub-groups \citep[see][Sect.~4 and 5]{Camera:2011ms}: the former collecting the parameters of interest, and the latter all those deemed `nuisance' or `systematics' for the analysis carried out. In the present case, the two subsets respectively refer to \textit{cosmological} and \textit{astrophysical} parameters. If there are $N_{\rm cosmo}$ cosmological parameters and $N_{\rm astro}$ astrophysical parameters, and indexes $\mu,\nu$ in \autoref{eq:Fisher} run from $1$ to $N_{\rm cosmo}+N_{\rm astro}$, we introduce two new sets of indices: $a,b=1\ldots N_{\rm cosmo}$, and $\alpha=1\ldots N_{\rm astro}$.

From this, the bias on a given cosmological parameter $p_a$, denoted by $b_{p_a}$, is computed via
\begin{equation}
b_{p_a} = \delta p_\alpha F_{b\alpha}\left(\mathbfss F^{-1}\right)_{ab}.\label{eq:bias}
\end{equation}
Here: $(\mathbfss F^{-1})_{ab}$ means taking the inverse of the full `cosmo'+`astro' Fisher matrix, and then selecting the elements relative to cosmological parameters only; $F_{b\alpha}$ are the elements of the rectangular sub-matrix mixing cosmological and astrophysical parameters; and $\delta p_\alpha$ is the vector of the shifts between the value of the astrophysical parameters assumed as fiducial and their true values, viz.\
\begin{equation}
\delta p_\alpha=p_\alpha^{\rm fid}-p_\alpha^{\rm true}.\label{eq:shift}
\end{equation}
Note that the Fisher matrix formalism is valid under the assumption that the likelihood function for the model parameters is well represented by a Gaussian in a neighbourhood of its peak. This means that the bias of \autoref{eq:bias} is valid over the whole parameter space if and only if the likelihood is exactly Gaussian. In other cases, these results are robust only up to a few standard deviations away from the fiducial values of the parameter set.

\begin{figure*}
\centering
\includegraphics[width=\textwidth]{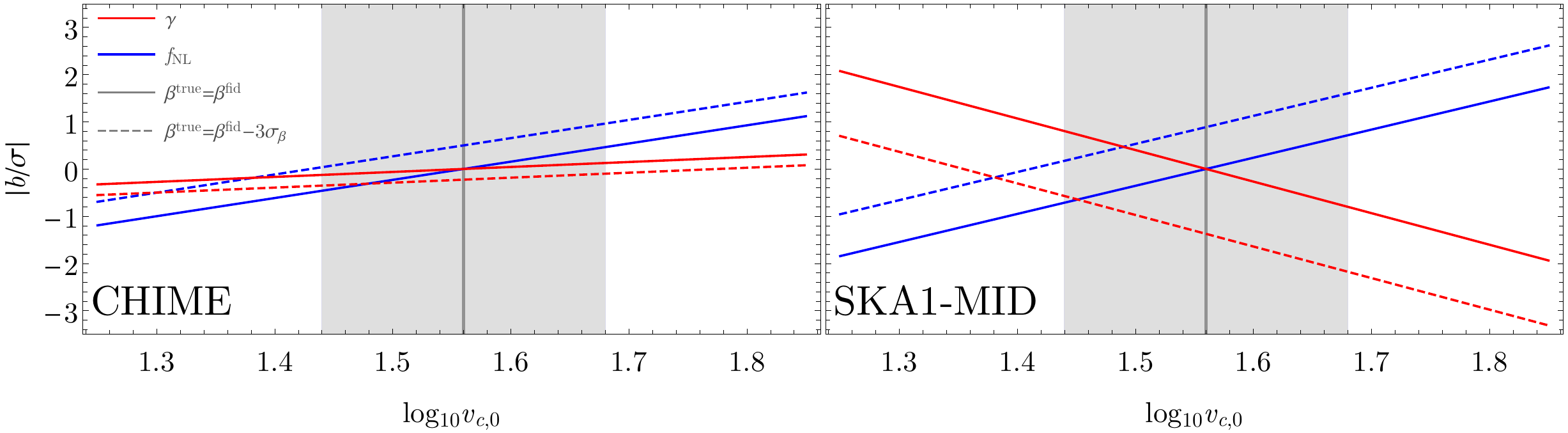}
\caption{Relative bias on modified gravity parameter, $\gamma$ (red curves), and primordial non-Gaussianity parameter, $\fnl$ (blue curves), arising from an incorrect assumption on the \hi\ astrophysical parameter $v_{{\rm c},0}$. The $x$-axis covers a wide range of the prior used for fitting the astrophysical parameters in \citet{Padmanabhan2017}, whose best-fit values (assumed here as fiducial) are indicated by the vertical line, whereas the grey-shaded areas mark the $\pm3\sigma$ confidence region around it found by \citet{Padmanabhan2017}. Whilst one of the astrophysical parameters, $v_{{\rm c},0}$, is left free to vary, the other, $\beta$, is kept fixed at its fiducial value (solid lines), or at the lower edge of its $3\sigma$ confidence interval (dashed lines). The experimental configurations adopted are that of CHIME (left panel) and SKA1-MID (right panel).}
\label{fig:relbias_vc0}
\end{figure*}

\begin{figure*}
\centering
\includegraphics[width=\textwidth]{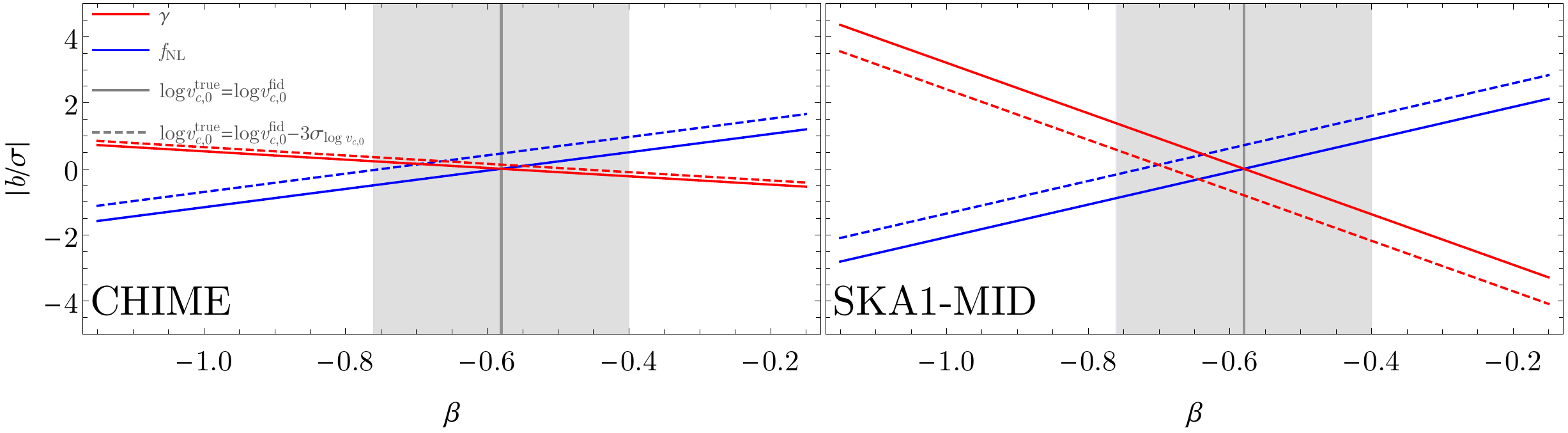}
\caption{Same as \autoref{fig:relbias_vc0}, but with $\beta$ free to vary and $v_{{\rm c},0}$ fixed at two different values (the best fit and $3\sigma$ bound edge, as found in \citealt{Padmanabhan2017}). Note the different extension of the $y$-axis, due to the difference in sensitivity to variations of the $\beta$ parameter.}
\label{fig:relbias_beta}
\end{figure*}

\section{Results and Discussion}
\label{sec:results}
The discussion below focusses on CHIME and SKA1-MID, as proxies of the interferometric and single dish experiments, respectively. Results for MeerKAT can be found in Appendix \ref{sec:extraexp}.

\subsection{Constraints on modified gravity and primordial non-Gaussianity}

In \autoref{fig:relbias_vc0} and \autoref{fig:relbias_beta} are plotted the relative bias on the beyond-\lcdm\ parameters $\gamma$ and $\fnl$, that arise from an incorrect assumption on the astrophysical parameters $v_{\rm c,0}$ and $\beta$ respectively. This represents how much the best-fit value of $\gamma$ and $\fnl$ would be wrongly estimated (in units of its standard deviation) if the true value of the astrophysical parameters were different from the fiducial ones assumed in the analysis (and indicated in \autoref{table:parameters}).

In each plot, the value of $v_{{\rm c},0}$ or $\beta$ respectively, is kept fixed and the value of the relative bias is plotted as a function of the other parameter, with red or blue curves for $\gamma$ or $\fnl$. The solid lines indicate the fixed parameter being kept at its fiducial value, which is also illustrated by the fact that the lines tend to zero bias at the fiducial values, $1.56$ for $\log_{10}v_{{\rm c},0}$ and $-0.58$ for $\beta$. Dashed lines show the cases of the fixed parameter being discrepant at $3\sigma$ from the best-fitting value found in \citet{Padmanabhan2017}. Note that \autoref{eq:shift} is symmetric in $\delta p_\alpha$, meaning that the parameter fiducial value being over- or under-estimated with respect to its true value only mirrors the curves with respect to the vertical, dark-grey line. The $3\sigma$ bounds on the free parameter in each case are indicated by the grey shaded region for clarity.

In a nutshell, the figures show that if the estimation of the astrophysical parameters is not too far off the fiducial, a measurement of $\fnl$ is rather robust with respect to the astrophysics. In other words, if both $\log _{10} v_{{\rm c},0}$ and $\beta$ are truly in a neighbourhood of $1.56$ and $-0.58$, respectively, the estimate of $\fnl$ will be biased by less than $2\sigma$ (in most cases, much less that $1\sigma$). Also, the figures indicate that the extended model parameters are more degenerate with larger values of $\beta$ than $v_{{\rm c},0}$.

For the modified gravity parameter $\gamma$, the biases can approach larger amounts when the astrophysical parameters are varied around $1$ to $3\sigma$ of their best-fit values; the numbers, however, remain within $2\sigma - 2.5\sigma$ in most cases.

\subsection{Extending the multipole range}
Thus far, the most part of the redshift bins considered
dealt with linear scales, where the \hi\ halo model may not play a very significant role. \citet{padmanabhan2019} explored extending the $\ell$-range up to $2000$, and a factor of a few improvement in the precision of cosmological and astrophysical constraints was found in that paper. In order to better assess the impact of the astrophysical modelling on the relative biases in parameter estimation, on even more nonlinear scales, we explore extending the $\ell$-range to about 4 times the fiducial. Specifically, we use 18 equi-log spaced bins with $\Delta(\log_{10} \ell_m) = 3/14$, starting again from $\ell=1$ but discarding the multipole bins that involve angular scales below the quadrupole.

\begin{figure*}
\centering
\includegraphics[width = \textwidth]{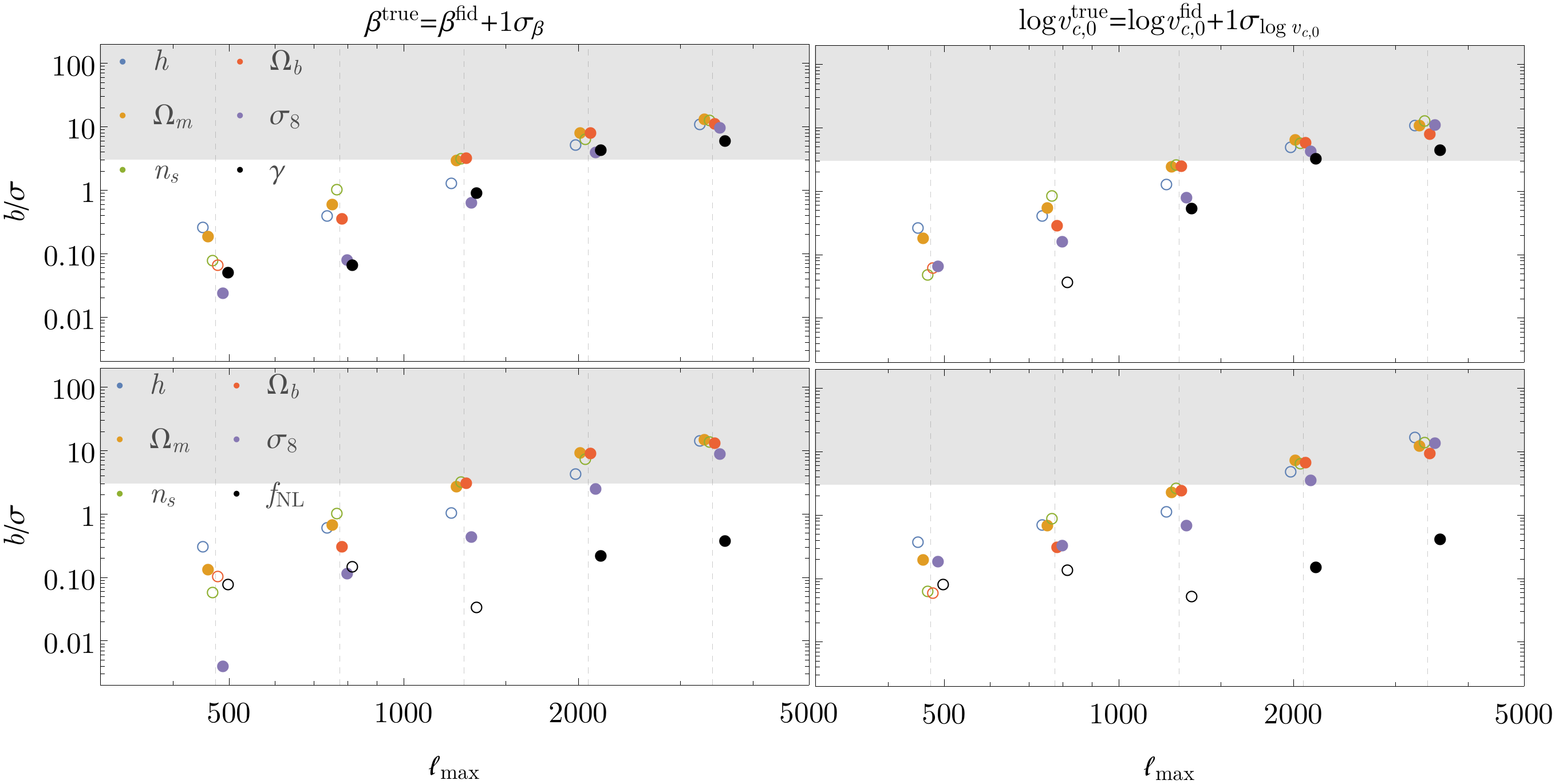}
\caption{Relative bias on the all the cosmological parameters as a function of the maximum multipole used in the analysis, obtained on shifting either astrophysical parameter, $\beta$ (left panels) or $v_{\rm c,0}$ (right panels), by $1 \sigma$ from its fiducial value. Top panels refer to the parameters in \lcdm+$\gamma$, whereas bottom panels refer to those in \lcdm+$\fnl$. Empty (filled) circles indicate negative (positive) values of the biases. Dashed vertical lines indicate the centres of the last five multipole bins of the extended $\ell$-range. The grey-shaded area marks relative bias values for which $b/\sigma>3$, for which results from the Fisher matrix bias formalism need to be taken with a grain of salt. The experimental configuration considered is that of CHIME.}
\label{fig:relbias1}
\end{figure*}
\begin{figure*}
\centering
\includegraphics[width=\textwidth]{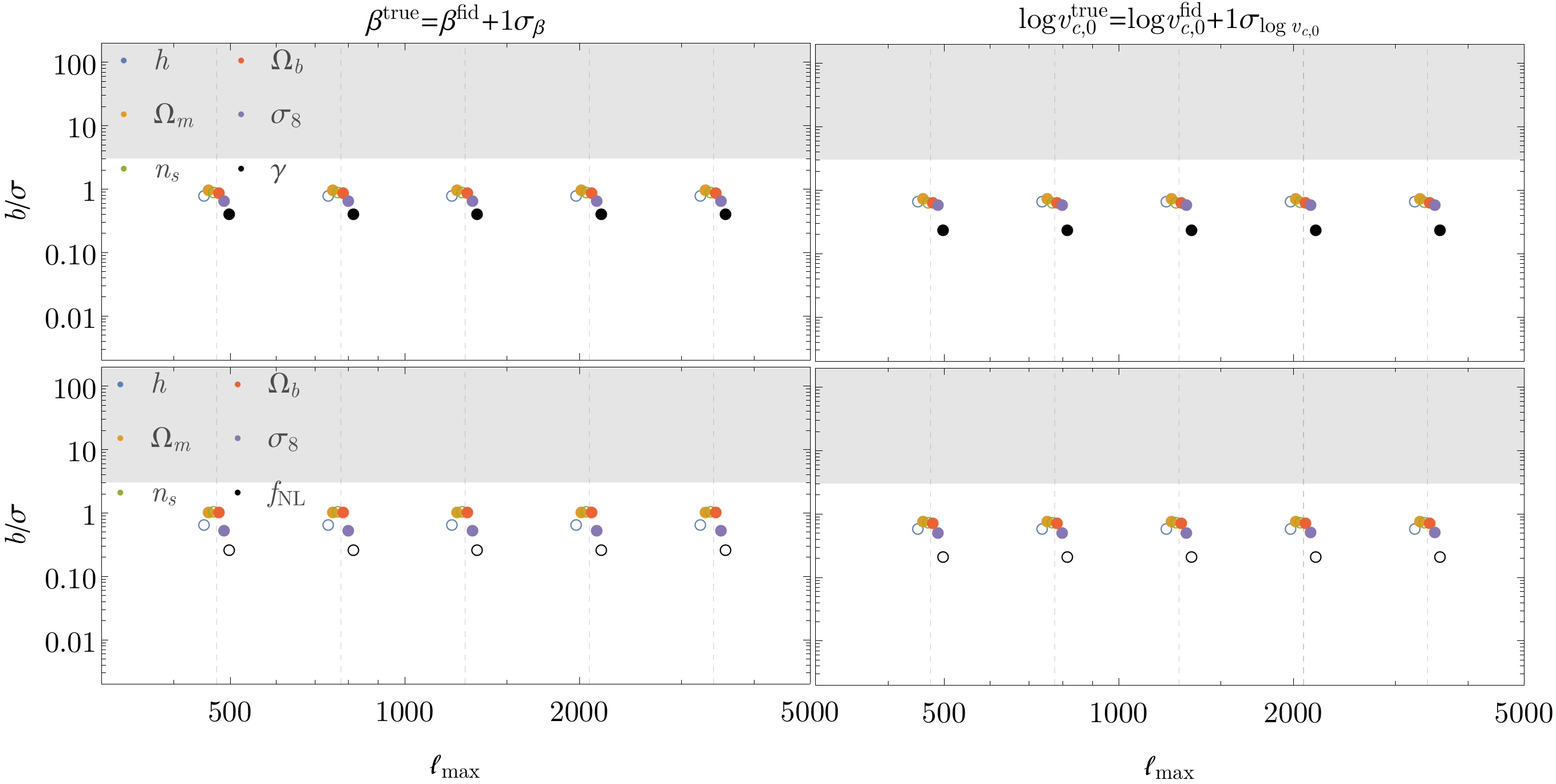}
\caption{Same as \autoref{fig:relbias1}, but for the case of SKA1-MID.}
\label{fig:relbias3}
\end{figure*}

In Figures \ref{fig:relbias1} and \ref{fig:relbias3}  are plotted the relative biases on all the cosmological parameters induced by an incorrect assumption on either astrophysical parameter, $v_{\rm c,0}$ and $\beta$ (each assumed to be shifted by $1\sigma$ from its fiducial value) for the CHIME-like and SKA I MID-like experimental configurations respectively. The relative bias is plotted as a function of the maximum multipole included in the analysis, with dashed vertical lines marking the centres of the last five $\ell$-bins. Filled circles are positive biases, whilst empty circles refer to negative ones. It is clear from these results that the interferometric set up of CHIME is well suited to explore the deeply nonlinear scales. In contrast, SKA1-MID is dominated by beam effects on the multipoles of the extended range considered in this subsection (the same is true for the MeerKAT configuration plotted in  \autoref{fig:relbias2} in the Appendix).

\begin{figure}
    \centering
    \includegraphics[width=\columnwidth]{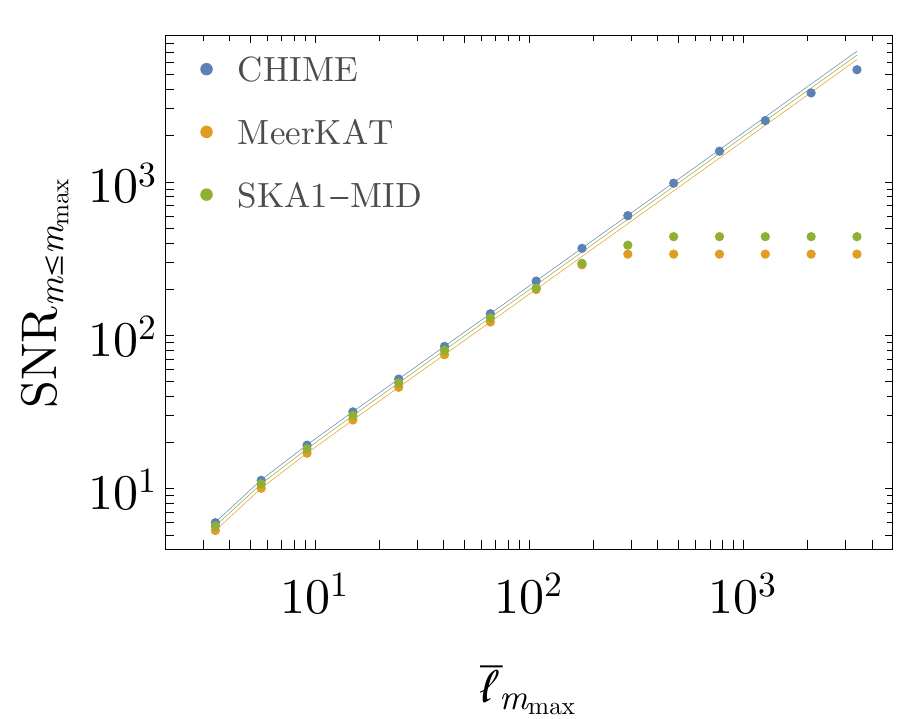}
    \caption{Cumulative SNR per multipole bin as a function of the highest multipole bin included. The different colours denote the three experimental configurations. Thin solid lines show, for comparison, the cosmic-variance limit for each configuration.}
    \label{fig:SNR_m}
\end{figure}

This last point can be better understood by looking at \autoref{fig:SNR_m}, which shows the cumulative signal-to-noise ratio (SNR) per multipole bin,
\begin{equation}
    {\rm SNR}_{m\le m_{\rm max}}=\sqrt{\sum_{i=1}^{N_z}\sum_{m=3}^{m_{\rm max}} \Delta\ell_m \left[\frac{C_{\bar{\ell}_m}^{\rm HI}(z_i)}{\Delta C_{\bar{\ell}_m}^{\rm HI}(z_i)}\right]^2},
    \label{eq:SNR_m}
\end{equation}
as a function of the maximum multipole included in its computation. The colour code, marking the different experiments, is the same as in \autoref{fig:Cl_comparison}. Here, each coloured circle denotes the cumulative SNR up to the angular scale in the $x$-axis corresponding to the maximum multipole bin. It is easy to see how for dish surveys the SNR flattens out as soon as the beam becomes dominant, whereas for an interferometer like CHIME, one keeps gaining SNR by including smaller and smaller scales. For comparison, the thin, solid lines show the cosmic-variance limit for each experiment. For instance, CHIME is cosmic-variance limited up to $\ell\approx700$, while both MeerKAT and SKA1-MID start feeling the thermal noise contribution at about $\ell \sim 100$.

 The noteworthy points emerging from Figures \ref{fig:relbias1} and \ref{fig:relbias3} are as follows:
\begin{itemize}
\item For all the standard cosmological parameters and the growth parameter (i.e.\ except for $\fnl$), the relative biases are within a few standard deviations---meaning they are acceptable---as long as $\ell_{\rm max} \lesssim 1500$; if we go to smaller scales, the uncertainties in the modelling of \hi\ clustering makes the cosmological parameter estimation unreliable. This is consistent with expectations, and provides a quantitative illustration of the regime in which the cosmological parameters are accurately recovered in the presence of astrophysical uncertainties.
\item Despite being dependent on the Gaussian halo bias, measurements of $\fnl$ are much more stable with respect to our knowledge of the \hi\ astrophysics, even if we go to very small scales.
\end{itemize}
These results essentially reconfirm the fact, also found previously, that $\fnl$ measurements probe the very largest scales, and hence are not `coupled' as strongly to the details of the \hi\ astrophysics compared to other cosmological parameters (including the modified gravity parameter $\gamma$). Remarkably, this continues to hold even though the $f_{\rm NL}$ parameter is strongly linked to the \hi\ bias (see \autoref{eq:delta_b}).

\section{Conclusions}
\label{sec:conclusions}
In this paper, we have investigated the dependence of the recovery of cosmological information from \hi\ intensity mapping surveys, on the details of the astrophysics of \hi. This is complementary to the study in \citet{padmanabhan2019}, which focusses on how much the accuracy (represented by the relative Fisher bias) of recovery of the parameters changes if the astrophysical parameters are allowed to deviate from the fiducial values constrained by all the observations available today \citep{Padmanabhan2017}. We have considered two effects beyond standard \lcdm\ cosmology: incorporating a primordial non-Gaussianity parameter $\fnl$, and exploring modifications to the growth parameter $\gamma$ which may result from deviations to general relativity.

We find that, although the $\fnl$ parameter heavily influences the halo bias (see \autoref{eq:delta_b}), the constraints on this parameter are considerably robust to astrophysical uncertainties, even at fairly nonlinear scales. Given that $\fnl$ is one of the strongest science cases for \hi\ intensity mapping, this result indicates that one can robustly use \hi\ intensity mapping to probe the largest cosmological scales in searches for primordial non-Gaussianity. 

We have also explored to what extent astrophysical uncertainties influence constraints on modified gravity models, taking as an illustration of this effect, the deviation of the growth parameter $\gamma$ from its fiducial value of $0.55$. It is important to note, however, that allowing $\gamma$ to deviate from its fiducial value may have broader consequences, especially in the context of the backreaction and averaging problems \citep[for a review, see, e.g.][]{clarkson2011}. Hence, constraints on $\gamma$ from future intensity mapping surveys acquire prime importance in the interpretation of these effects as well. Unlike the primordial non-Gaussianity parameter, we find that growth parameter is more `connected' to the astrophysical uncertainties, in a similar manner as the standard cosmological parameters. For both, the standard cosmological parameters and the growth parameter, the biases in accuracy can be mitigated either by neglecting information coming from nonlinear scales (where astrophysics plays a crucial role), or by jointly fitting for cosmological \textit{and} astrophysical parameters at the same time.

As mentioned previously in Sec. \ref{sec:halomodel}, the framework used for populating neutral hydrogen into dark matter haloes has its basis in the combination of the presently available data. The framework lends itself naturally to a Fisher matrix procedure, facilitates the addition of new physics, and allows the comparison of cosmological and astrophysical parameters at an equivalent level. Moreover, the resultant prior range of the astrophysical parameters becomes meaningful only in this framework, since the uncertainties on the astrophysical parameters represent our currently available best constraints. 
However, in the light of future data from upcoming experiments, it may become possible to constrain a more extended \hi-halo mass relation. This possibility was considered in \citet{padmanabhan2019}, in which the effect of an extended astrophysical parameterisation on the recovery of $\Lambda$CDM cosmology was explored in detail. In so doing, it was found that the cosmological forecasts are negligibly degraded by the addition of a reasonable number of new parameters extending the existing framework. This finding is expected to generalise to the case of the present analysis as well, when we note that the cosmological parameters $f_{\rm NL}$ and $\gamma$ are less strongly coupled to the astrophysics than those describing the standard $\Lambda$CDM. Thus, we can conclude that the cosmological recovery from \hi\ intensity mapping is in general not expected to be sensitive to the choice of astrophysical parameterisation used in the analysis. 

It is also worth commenting upon the degeneracy structure of the astrophysical and cosmological parameters when the halo model framework is constrained by the angular power spectrum. It is found (e.g. Fig.~2 of \citealt{padmanabhan2019}) that the parameters $\sigma_8$, $n_{\rm s}$ and $h$ are the most sensitive to changes in the astrophysical parameters $v_{\rm c,0} $ and $\beta$, while the abundance parameters $\om$ and $\ob$ are less affected. This arises due to the former, like $\gamma$ and $f_{\rm NL}$, being closely linked to $P_{\rm HI} (k,z)$, while the latter are connected to the overall normalisation of $M_{\rm HI} (M,z)$ and thus to the astrophysical $\alpha$ parameter, which, as we have seen before, is not constrainable from a dimensionless power spectrum.

In general, measurements of primordial non-Gaussianity acquire significant biases when general relativistic effects are not taken into account \citep{Camera:2014sba}. However, the main contributor to this systematic bias is lensing magnification \citep{Alonso:2015sfa}, an effect which is not present in intensity mapping \citep{Hall:2012wd} as a consequence of surface brightness conservation. Hence, we expect general relativistic corrections to have negligible effects on our constraints on the $\fnl$ parameter. Our results for $\fnl$ are generally applicable to cases of cosmological modifications which probe scales `disconnected' from astrophysical uncertainties.

At present, we have neglected the effects of foregrounds, which are expected to be the limiting systematic in these surveys. However, these systematics will more heavily affect the largest scales, implying no loss of generality concerning our results. Multi-tracer methods, on the other hand, offer salient avenues to mitigate the effects of uncorrelated foregrounds \citep[e.g.][]{cunnington2019,modi2019}. We hope to address these and other horizon-scale effects constrained by intensity mapping in future work.

\section*{Acknowledgements}
We thank the referee for a detailed and helpful report that improved the content and presentation. SC is supported by the Italian Ministry of Education, University and Research (\textsc{miur}) through Rita Levi Montalcini project `\textsc{prometheus} -- Probing and Relating Observables with Multi-wavelength Experiments To Help Enlightening the Universe's Structure', and by the `Departments of Excellence 2018-2022' Grant awarded by \textsc{miur} (L.\ 232/2016). 




\bibliographystyle{mnras}
\bibliography{main_bib} 

\begin{thebibliography}{}
\makeatletter
\relax
\def\mn@urlcharsother{\let\do\@makeother \do\$\do\&\do\#\do\^\do\_\do\%\do\~}
\def\mn@doi{\begingroup\mn@urlcharsother \@ifnextchar [ {\mn@doi@}
  {\mn@doi@[]}}
\def\mn@doi@[#1]#2{\def\@tempa{#1}\ifx\@tempa\@empty \href
  {http://dx.doi.org/#2} {doi:#2}\else \href {http://dx.doi.org/#2} {#1}\fi
  \endgroup}
\def\mn@eprint#1#2{\mn@eprint@#1:#2::\@nil}
\def\mn@eprint@arXiv#1{\href {http://arxiv.org/abs/#1} {{\tt arXiv:#1}}}
\def\mn@eprint@dblp#1{\href {http://dblp.uni-trier.de/rec/bibtex/#1.xml}
  {dblp:#1}}
\def\mn@eprint@#1:#2:#3:#4\@nil{\def\@tempa {#1}\def\@tempb {#2}\def\@tempc
  {#3}\ifx \@tempc \@empty \let \@tempc \@tempb \let \@tempb \@tempa \fi \ifx
  \@tempb \@empty \def\@tempb {arXiv}\fi \@ifundefined
  {mn@eprint@\@tempb}{\@tempb:\@tempc}{\expandafter \expandafter \csname
  mn@eprint@\@tempb\endcsname \expandafter{\@tempc}}}

\bibitem[\protect\citeauthoryear{Ade et~al.}{Ade et~al.}{2016}]{Ade:2015xua}
Ade P. A.~R.,  et~al., 2016, \mn@doi [Astron. Astrophys.]
  {10.1051/0004-6361/201525830}, 594, A13

\bibitem[\protect\citeauthoryear{Alonso \& Ferreira}{Alonso \&
  Ferreira}{2015}]{Alonso:2015sfa}
Alonso D.,  Ferreira P.~G.,  2015, \mn@doi [Phys. Rev.]
  {10.1103/PhysRevD.92.063525}, D92, 063525

\bibitem[\protect\citeauthoryear{Alonso, Bull, Ferreira, Maartens  \&
  Santos}{Alonso et~al.}{2015}]{Alonso:2015uua}
Alonso D.,  Bull P.,  Ferreira P.~G.,  Maartens R.,   Santos M.,  2015, \mn@doi
  [Astrophys. J.] {10.1088/0004-637X/814/2/145}, 814, 145

\bibitem[\protect\citeauthoryear{Amendola et~al.}{Amendola
  et~al.}{2013}]{Amendola:2012ys}
Amendola L.,  et~al., 2013, Living Rev. Rel., 16, 6

\bibitem[\protect\citeauthoryear{Amendola et~al.}{Amendola
  et~al.}{2018}]{Amendola:2016saw}
Amendola L.,  et~al., 2018, \mn@doi [Living Rev. Rel.]
  {10.1007/s41114-017-0010-3}, 21, 2

\bibitem[\protect\citeauthoryear{{Anderson} et~al.,}{{Anderson}
  et~al.}{2018}]{anderson2018}
{Anderson} C.~J.,  et~al., 2018, \mn@doi [\mnras] {10.1093/mnras/sty346}, \href
  {https://ui.adsabs.harvard.edu/\#abs/2018MNRAS.476.3382A} {476, 3382}

\bibitem[\protect\citeauthoryear{{Baker} \& {Bull}}{{Baker} \&
  {Bull}}{2015}]{2015ApJ...811..116B}
{Baker} T.,  {Bull} P.,  2015, \mn@doi [\apj] {10.1088/0004-637X/811/2/116},
  \href {https://ui.adsabs.harvard.edu/abs/2015ApJ...811..116B} {811, 116}

\bibitem[\protect\citeauthoryear{{Ballardini} \& {Maartens}}{{Ballardini} \&
  {Maartens}}{2019}]{ballardini2019}
{Ballardini} M.,  {Maartens} R.,  2019, \mn@doi [\mnras]
  {10.1093/mnras/stz480}, \href
  {https://ui.adsabs.harvard.edu/abs/2019MNRAS.485.1339B} {485, 1339}

\bibitem[\protect\citeauthoryear{{Ballardini}, {Matthewson}  \&
  {Maartens}}{{Ballardini} et~al.}{2019}]{ballardini2019mt}
{Ballardini} M.,  {Matthewson} W.~L.,   {Maartens} R.,  2019, \mn@doi [\mnras]
  {10.1093/mnras/stz2258}, \href
  {https://ui.adsabs.harvard.edu/abs/2019MNRAS.489.1950B} {489, 1950}

\bibitem[\protect\citeauthoryear{Bartolo, Komatsu, Matarrese  \&
  Riotto}{Bartolo et~al.}{2004}]{Bartolo:2004if}
Bartolo N.,  Komatsu E.,  Matarrese S.,   Riotto A.,  2004, \mn@doi [Phys.
  Rept.] {10.1016/j.physrep.2004.08.022}, 402, 103

\bibitem[\protect\citeauthoryear{{Bernal}, {Breysse}, {Gil-Mar{\'\i}n}  \&
  {Kovetz}}{{Bernal} et~al.}{2019}]{bernal2019}
{Bernal} J.~L.,  {Breysse} P.~C.,  {Gil-Mar{\'\i}n} H.,   {Kovetz} E.~D.,
  2019, arXiv e-prints, \href
  {https://ui.adsabs.harvard.edu/abs/2019arXiv190710067B} {p. arXiv:1907.10067}

\bibitem[\protect\citeauthoryear{{Bull}, {Ferreira}, {Patel}  \&
  {Santos}}{{Bull} et~al.}{2014a}]{bull2014}
{Bull} P.,  {Ferreira} P.~G.,  {Patel} P.,   {Santos} M.~G.,  2014a,
  arXiv:1405.1452, \href {http://adsabs.harvard.edu/abs/2014arXiv1405.1452B} {}

\bibitem[\protect\citeauthoryear{Bull, Ferreira, Patel  \& Santos}{Bull
  et~al.}{2014b}]{Bull:2014rha}
Bull P.,  Ferreira P.~G.,  Patel P.,   Santos M.~G.,  2014b, eprint
  arXiv:1405.1452

\bibitem[\protect\citeauthoryear{Bull et~al.}{Bull et~al.}{2016}]{Bull:2015stt}
Bull P.,  et~al., 2016, \mn@doi [Phys. Dark Univ.]
  {10.1016/j.dark.2016.02.001}, 12, 56

\bibitem[\protect\citeauthoryear{Camera, Diaferio  \& Cardone}{Camera
  et~al.}{2011}]{Camera:2011ms}
Camera S.,  Diaferio A.,   Cardone V.~F.,  2011, \mn@doi [JCAP]
  {10.1088/1475-7516/2011/07/016}, 1107, 016

\bibitem[\protect\citeauthoryear{Camera, Santos, Ferreira  \&
  Ferramacho}{Camera et~al.}{2013}]{Camera:2013kpa}
Camera S.,  Santos M.~G.,  Ferreira P.~G.,   Ferramacho L.,  2013, \mn@doi
  [Phys. Rev. Lett.] {10.1103/PhysRevLett.111.171302}, 111, 171302

\bibitem[\protect\citeauthoryear{Camera, Maartens  \& Santos}{Camera
  et~al.}{2015a}]{Camera:2014sba}
Camera S.,  Maartens R.,   Santos M.~G.,  2015a, \mn@doi [Mon. Not. Roy.
  Astron. Soc.] {10.1093/mnrasl/slv069}, 451, L80

\bibitem[\protect\citeauthoryear{Camera, Raccanelli, Bull, Bertacca, Chen
  et~al.}{Camera et~al.}{2015b}]{Camera:2015fsa}
Camera S.,  Raccanelli A.,  Bull P.,  Bertacca D.,  Chen X.,   et~al., 2015b,
  PoS, AASKA14, 025

\bibitem[\protect\citeauthoryear{Camera, Harrison, Bonaldi  \& Brown}{Camera
  et~al.}{2017}]{Camera:2016owj}
Camera S.,  Harrison I.,  Bonaldi A.,   Brown M.~L.,  2017, \mn@doi [Mon. Not.
  Roy. Astron. Soc.] {10.1093/mnras/stw2688}, 464, 4747

\bibitem[\protect\citeauthoryear{Camera, Fonseca, Maartens  \& Santos}{Camera
  et~al.}{2018}]{Camera:2018jys}
Camera S.,  Fonseca J.,  Maartens R.,   Santos M.~G.,  2018, \mn@doi [Mon. Not.
  Roy. Astron. Soc.] {10.1093/mnras/sty2284}, 481, 1251

\bibitem[\protect\citeauthoryear{{Castorina} \& {White}}{{Castorina} \&
  {White}}{2019}]{castorina2019a}
{Castorina} E.,  {White} M.,  2019, \mn@doi [\jcap]
  {10.1088/1475-7516/2019/06/025}, \href
  {https://ui.adsabs.harvard.edu/abs/2019JCAP...06..025C} {2019, 025}

\bibitem[\protect\citeauthoryear{{Castorina} et~al.,}{{Castorina}
  et~al.}{2019}]{castorina2019}
{Castorina} E.,  et~al., 2019, arXiv e-prints, \href
  {https://ui.adsabs.harvard.edu/abs/2019arXiv190408859C} {p. arXiv:1904.08859}

\bibitem[\protect\citeauthoryear{{Chang}, {Pen}, {Bandura}  \&
  {Peterson}}{{Chang} et~al.}{2010}]{chang10}
{Chang} T.-C.,  {Pen} U.-L.,  {Bandura} K.,   {Peterson} J.~B.,  2010, \mn@doi
  [\nat] {10.1038/nature09187}, \href
  {http://adsabs.harvard.edu/abs/2010Natur.466..463C} {466, 463}

\bibitem[\protect\citeauthoryear{{Clarkson}, {Ellis}, {Larena}  \&
  {Umeh}}{{Clarkson} et~al.}{2011}]{clarkson2011}
{Clarkson} C.,  {Ellis} G.,  {Larena} J.,   {Umeh} O.,  2011, arXiv e-prints,
  \href {https://ui.adsabs.harvard.edu/abs/2011arXiv1109.2314C} {p.
  arXiv:1109.2314}

\bibitem[\protect\citeauthoryear{Clifton, Ferreira, Padilla  \&
  Skordis}{Clifton et~al.}{2012}]{Clifton:2011jh}
Clifton T.,  Ferreira P.~G.,  Padilla A.,   Skordis C.,  2012, \mn@doi [Phys.
  Rept.] {10.1016/j.physrep.2012.01.001}, 513, 1

\bibitem[\protect\citeauthoryear{{Cosmic Visions 21 cm Collaboration}
  et~al.,}{{Cosmic Visions 21 cm Collaboration}
  et~al.}{2018}]{cosmicvisions2018}
{Cosmic Visions 21 cm Collaboration} et~al., 2018, arXiv e-prints, \href
  {https://ui.adsabs.harvard.edu/\#abs/2018arXiv181009572C} {p.
  arXiv:1810.09572}

\bibitem[\protect\citeauthoryear{{Cunnington}, {Wolz}, {Pourtsidou}  \&
  {Bacon}}{{Cunnington} et~al.}{2019}]{cunnington2019}
{Cunnington} S.,  {Wolz} L.,  {Pourtsidou} A.,   {Bacon} D.,  2019, arXiv
  e-prints, \href {https://ui.adsabs.harvard.edu/abs/2019arXiv190401479C} {p.
  arXiv:1904.01479}

\bibitem[\protect\citeauthoryear{Dalal, Dore, Huterer  \& Shirokov}{Dalal
  et~al.}{2008}]{Dalal:2007cu}
Dalal N.,  Dore O.,  Huterer D.,   Shirokov A.,  2008, \mn@doi [Phys. Rev.]
  {10.1103/PhysRevD.77.123514}, D77, 123514

\bibitem[\protect\citeauthoryear{{Euclid Collaboration} et~al.,}{{Euclid
  Collaboration} et~al.}{2019}]{2019arXiv191009273E}
{Euclid Collaboration} et~al., 2019, arXiv e-prints, \href
  {https://ui.adsabs.harvard.edu/abs/2019arXiv191009273E} {p. arXiv:1910.09273}

\bibitem[\protect\citeauthoryear{Fonseca, Camera, Santos  \& Maartens}{Fonseca
  et~al.}{2015}]{Fonseca:2015laa}
Fonseca J.,  Camera S.,  Santos M.,   Maartens R.,  2015, \mn@doi [Astrophys.
  J.] {10.1088/2041-8205/812/2/L22}, 812, L22

\bibitem[\protect\citeauthoryear{Fonseca, Maartens  \& Santos}{Fonseca
  et~al.}{2017}]{Fonseca:2016xvi}
Fonseca J.,  Maartens R.,   Santos M.~G.,  2017, \mn@doi [Mon. Not. Roy.
  Astron. Soc.] {10.1093/mnras/stw3248}, 466, 2780

\bibitem[\protect\citeauthoryear{{Fonseca}, {Viljoen}  \& {Maartens}}{{Fonseca}
  et~al.}{2019}]{fonseca2019}
{Fonseca} J.,  {Viljoen} J.-A.,   {Maartens} R.,  2019, arXiv e-prints, \href
  {https://ui.adsabs.harvard.edu/abs/2019arXiv190702975F} {p. arXiv:1907.02975}

\bibitem[\protect\citeauthoryear{{Font-Ribera} et~al.,}{{Font-Ribera}
  et~al.}{2012}]{fontribera2012}
{Font-Ribera} A.,  et~al., 2012, \mn@doi [\jcap]
  {10.1088/1475-7516/2012/11/059}, \href
  {http://adsabs.harvard.edu/abs/2012JCAP...11..059F} {11, 59}

\bibitem[\protect\citeauthoryear{{Giannantonio}, {Ross}, {Percival},
  {Crittenden}, {Bacher}, {Kilbinger}, {Nichol}  \& {Weller}}{{Giannantonio}
  et~al.}{2014}]{2014PhRvD..89b3511G}
{Giannantonio} T.,  {Ross} A.~J.,  {Percival} W.~J.,  {Crittenden} R.,
  {Bacher} D.,  {Kilbinger} M.,  {Nichol} R.,   {Weller} J.,  2014, \mn@doi
  [\prd] {10.1103/PhysRevD.89.023511}, \href
  {https://ui.adsabs.harvard.edu/abs/2014PhRvD..89b3511G} {89, 023511}

\bibitem[\protect\citeauthoryear{Hall, Bonvin  \& Challinor}{Hall
  et~al.}{2013}]{Hall:2012wd}
Hall A.,  Bonvin C.,   Challinor A.,  2013, \mn@doi [Phys. Rev.]
  {10.1103/PhysRevD.87.064026}, D87, 064026

\bibitem[\protect\citeauthoryear{Heavens, Kitching  \& Verde}{Heavens
  et~al.}{2007}]{Heavens:2007ka}
Heavens A.~F.,  Kitching T.~D.,   Verde L.,  2007, \mn@doi [Mon. Not. Roy.
  Astron. Soc.] {10.1111/j.1365-2966.2007.12134.x}, 380, 1029

\bibitem[\protect\citeauthoryear{Jalilvand, Ghosh, Majerotto, Bose, Durrer  \&
  Kunz}{Jalilvand et~al.}{2019a}]{Jalilvand:2019brk}
Jalilvand M.,  Ghosh B.,  Majerotto E.,  Bose B.,  Durrer R.,   Kunz M.,  2019a

\bibitem[\protect\citeauthoryear{{Jalilvand}, {Majerotto}, {Durrer}  \&
  {Kunz}}{{Jalilvand} et~al.}{2019b}]{jalilvand2019}
{Jalilvand} M.,  {Majerotto} E.,  {Durrer} R.,   {Kunz} M.,  2019b, \mn@doi
  [\jcap] {10.1088/1475-7516/2019/01/020}, \href
  {https://ui.adsabs.harvard.edu/abs/2019JCAP...01..020J} {2019, 020}

\bibitem[\protect\citeauthoryear{Knox}{Knox}{1995}]{Knox:1995dq}
Knox L.,  1995, \mn@doi [Phys. Rev.] {10.1103/PhysRevD.52.4307}, D52, 4307

\bibitem[\protect\citeauthoryear{Komatsu et~al.}{Komatsu
  et~al.}{2011}]{Komatsu:2010fb}
Komatsu E.,  et~al., 2011, \mn@doi [Astrophys. J. Suppl.]
  {10.1088/0067-0049/192/2/18}, 192, 18

\bibitem[\protect\citeauthoryear{{Kovetz} et~al.,}{{Kovetz}
  et~al.}{2019}]{kovetz2019}
{Kovetz} E.~D.,  et~al., 2019, arXiv e-prints, \href
  {https://ui.adsabs.harvard.edu/\#abs/2019arXiv190304496K} {p.
  arXiv:1903.04496}

\bibitem[\protect\citeauthoryear{Lahav, Lilje, Primack  \& Rees}{Lahav
  et~al.}{1991}]{Lahav:1991wc}
Lahav O.,  Lilje P.~B.,  Primack J.~R.,   Rees M.~J.,  1991, Mon. Not. Roy.
  Astron. Soc., 251, 128

\bibitem[\protect\citeauthoryear{Lewis, Challinor  \& Lasenby}{Lewis
  et~al.}{2000}]{Lewis:1999bs}
Lewis A.,  Challinor A.,   Lasenby A.,  2000, \mn@doi [Astrophys. J.]
  {10.1086/309179}, 538, 473

\bibitem[\protect\citeauthoryear{Linder}{Linder}{2005}]{Linder:2005in}
Linder E.~V.,  2005, \mn@doi [Phys. Rev.] {10.1103/PhysRevD.72.043529}, D72,
  043529

\bibitem[\protect\citeauthoryear{LoVerde \& Afshordi}{LoVerde \&
  Afshordi}{2008}]{LoVerde:2008re}
LoVerde M.,  Afshordi N.,  2008, \mn@doi [Phys. Rev.]
  {10.1103/PhysRevD.78.123506}, D78, 123506

\bibitem[\protect\citeauthoryear{{Loeb} \& {Wyithe}}{{Loeb} \&
  {Wyithe}}{2008}]{loeb2008}
{Loeb} A.,  {Wyithe} J. S.~B.,  2008, \mn@doi [\prl]
  {10.1103/PhysRevLett.100.161301}, \href
  {https://ui.adsabs.harvard.edu/abs/2008PhRvL.100p1301L} {100, 161301}

\bibitem[\protect\citeauthoryear{{Macci{\`o}}, {Dutton}, {van den Bosch},
  {Moore}, {Potter}  \& {Stadel}}{{Macci{\`o}} et~al.}{2007}]{maccio2007}
{Macci{\`o}} A.~V.,  {Dutton} A.~A.,  {van den Bosch} F.~C.,  {Moore} B.,
  {Potter} D.,   {Stadel} J.,  2007, \mn@doi [\mnras]
  {10.1111/j.1365-2966.2007.11720.x}, \href
  {http://adsabs.harvard.edu/abs/2007MNRAS.378...55M} {378, 55}

\bibitem[\protect\citeauthoryear{Maldacena}{Maldacena}{2003}]{Maldacena:2002vr}
Maldacena J.~M.,  2003, JHEP, 0305, 013

\bibitem[\protect\citeauthoryear{{Mana}, {Giannantonio}, {Weller}, {Hoyle},
  {H{\"u}tsi}  \& {Sartoris}}{{Mana} et~al.}{2013}]{2013MNRAS.434..684M}
{Mana} A.,  {Giannantonio} T.,  {Weller} J.,  {Hoyle} B.,  {H{\"u}tsi} G.,
  {Sartoris} B.,  2013, \mn@doi [\mnras] {10.1093/mnras/stt1062}, \href
  {https://ui.adsabs.harvard.edu/abs/2013MNRAS.434..684M} {434, 684}

\bibitem[\protect\citeauthoryear{{Martin}, {Giovanelli}, {Haynes}  \&
  {Guzzo}}{{Martin} et~al.}{2012}]{martin12}
{Martin} A.~M.,  {Giovanelli} R.,  {Haynes} M.~P.,   {Guzzo} L.,  2012, \mn@doi
  [\apj] {10.1088/0004-637X/750/1/38}, \href
  {http://adsabs.harvard.edu/abs/2012ApJ...750...38M} {750, 38}

\bibitem[\protect\citeauthoryear{{Masui}, {Schmidt}, {Pen}  \&
  {McDonald}}{{Masui} et~al.}{2010}]{masui2010}
{Masui} K.~W.,  {Schmidt} F.,  {Pen} U.-L.,   {McDonald} P.,  2010, \mn@doi
  [\prd] {10.1103/PhysRevD.81.062001}, \href
  {https://ui.adsabs.harvard.edu/abs/2010PhRvD..81f2001M} {81, 062001}

\bibitem[\protect\citeauthoryear{{Masui} et~al.,}{{Masui}
  et~al.}{2013}]{masui13}
{Masui} K.~W.,  et~al., 2013, \mn@doi [\apjl] {10.1088/2041-8205/763/1/L20},
  \href {http://adsabs.harvard.edu/abs/2013ApJ...763L..20M} {763, L20}

\bibitem[\protect\citeauthoryear{{Modi}, {White}, {Slosar}  \&
  {Castorina}}{{Modi} et~al.}{2019}]{modi2019}
{Modi} C.,  {White} M.,  {Slosar} A.,   {Castorina} E.,  2019, arXiv e-prints,
  \href {https://ui.adsabs.harvard.edu/abs/2019arXiv190702330M} {p.
  arXiv:1907.02330}

\bibitem[\protect\citeauthoryear{{Moradinezhad Dizgah}, {Keating}  \&
  {Fialkov}}{{Moradinezhad Dizgah} et~al.}{2019}]{dizgah2019}
{Moradinezhad Dizgah} A.,  {Keating} G.~K.,   {Fialkov} A.,  2019, \mn@doi [The
  Astrophysical Journal] {10.3847/2041-8213/aaf813}, \href
  {https://ui.adsabs.harvard.edu/abs/2019ApJ...870L...4M} {870, L4}

\bibitem[\protect\citeauthoryear{{Newburgh} et~al.,}{{Newburgh}
  et~al.}{2014}]{newburgh2014}
{Newburgh} L.~B.,  et~al., 2014, {Calibrating CHIME: a new radio interferometer
  to probe dark energy}.
p. 91454V, \mn@doi{10.1117/12.2056962}

\bibitem[\protect\citeauthoryear{{Noterdaeme} et~al.,}{{Noterdaeme}
  et~al.}{2012}]{noterdaeme12}
{Noterdaeme} P.,  et~al., 2012, \mn@doi [\aap] {10.1051/0004-6361/201220259},
  \href {http://adsabs.harvard.edu/abs/2012A%26A...547L...1N} {547, L1}

\bibitem[\protect\citeauthoryear{{Obuljen}, {Castorina}, {Villaescusa-Navarro}
  \& {Viel}}{{Obuljen} et~al.}{2018}]{obuljen2018}
{Obuljen} A.,  {Castorina} E.,  {Villaescusa-Navarro} F.,   {Viel} M.,  2018,
  \mn@doi [\jcap] {10.1088/1475-7516/2018/05/004}, \href
  {https://ui.adsabs.harvard.edu/abs/2018JCAP...05..004O} {2018, 004}

\bibitem[\protect\citeauthoryear{{Padmanabhan}, {Refregier}  \&
  {Amara}}{{Padmanabhan} et~al.}{2017}]{Padmanabhan2017}
{Padmanabhan} H.,  {Refregier} A.,   {Amara} A.,  2017, \mn@doi [\mnras]
  {10.1093/mnras/stx979}, \href
  {http://adsabs.harvard.edu/abs/2017MNRAS.469.2323P} {469, 2323}

\bibitem[\protect\citeauthoryear{{Padmanabhan}, {Refregier}  \&
  {Amara}}{{Padmanabhan} et~al.}{2019}]{padmanabhan2019}
{Padmanabhan} H.,  {Refregier} A.,   {Amara} A.,  2019, \mn@doi [Monthly
  Notices of the Royal Astronomical Society] {10.1093/mnras/stz683}, \href
  {https://ui.adsabs.harvard.edu/abs/2019MNRAS.485.4060P} {485, 4060}

\bibitem[\protect\citeauthoryear{{Planck Collaboration} et~al.,}{{Planck
  Collaboration} et~al.}{2018}]{planck2018}
{Planck Collaboration} et~al., 2018, arXiv e-prints, \href
  {https://ui.adsabs.harvard.edu/abs/2018arXiv180706209P} {p. arXiv:1807.06209}

\bibitem[\protect\citeauthoryear{{Planck Collaboration} et~al.,}{{Planck
  Collaboration} et~al.}{2019}]{planck2019}
{Planck Collaboration} et~al., 2019, arXiv e-prints, \href
  {https://ui.adsabs.harvard.edu/abs/2019arXiv190505697P} {p. arXiv:1905.05697}

\bibitem[\protect\citeauthoryear{{Pourtsidou}, {Bacon}, {Crittenden}  \&
  {Metcalf}}{{Pourtsidou} et~al.}{2016}]{pourtsidou2016}
{Pourtsidou} A.,  {Bacon} D.,  {Crittenden} R.,   {Metcalf} R.~B.,  2016,
  \mn@doi [\mnras] {10.1093/mnras/stw658}, \href
  {https://ui.adsabs.harvard.edu/\#abs/2016MNRAS.459..863P} {459, 863}

\bibitem[\protect\citeauthoryear{{Raccanelli} et~al.,}{{Raccanelli}
  et~al.}{2015}]{racanelli2015}
{Raccanelli} A.,  et~al., 2015, \mn@doi [Journal of Cosmology and
  Astro-Particle Physics] {10.1088/1475-7516/2015/01/042}, \href
  {https://ui.adsabs.harvard.edu/abs/2015JCAP...01..042R} {2015, 042}

\bibitem[\protect\citeauthoryear{{Rao}, {Turnshek}  \& {Nestor}}{{Rao}
  et~al.}{2006}]{rao06}
{Rao} S.~M.,  {Turnshek} D.~A.,   {Nestor} D.~B.,  2006, \mn@doi [\apj]
  {10.1086/498132}, \href {http://adsabs.harvard.edu/abs/2006ApJ...636..610R}
  {636, 610}

\bibitem[\protect\citeauthoryear{{Scoccimarro}, {Sheth}, {Hui}  \&
  {Jain}}{{Scoccimarro} et~al.}{2001}]{scoccimarro2001}
{Scoccimarro} R.,  {Sheth} R.~K.,  {Hui} L.,   {Jain} B.,  2001, \mn@doi [\apj]
  {10.1086/318261}, \href {http://adsabs.harvard.edu/abs/2001ApJ...546...20S}
  {546, 20}

\bibitem[\protect\citeauthoryear{{Seljak}}{{Seljak}}{2009}]{seljak2009}
{Seljak} U.,  2009, \mn@doi [\prl] {10.1103/PhysRevLett.102.021302}, \href
  {https://ui.adsabs.harvard.edu/\#abs/2009PhRvL.102b1302S} {102, 021302}

\bibitem[\protect\citeauthoryear{Sheth \& Tormen}{Sheth \&
  Tormen}{2002}]{Sheth:2001dp}
Sheth R.~K.,  Tormen G.,  2002, \mn@doi [Mon. Not. Roy. Astron. Soc.]
  {10.1046/j.1365-8711.2002.04950.x}, 329, 61

\bibitem[\protect\citeauthoryear{{Square Kilometre Array Cosmology Science
  Working Group} et~al.,}{{Square Kilometre Array Cosmology Science Working
  Group} et~al.}{2018}]{skacosmo2018}
{Square Kilometre Array Cosmology Science Working Group} et~al., 2018, arXiv
  e-prints, \href {https://ui.adsabs.harvard.edu/abs/2018arXiv181102743S} {p.
  arXiv:1811.02743}

\bibitem[\protect\citeauthoryear{{Switzer} et~al.,}{{Switzer}
  et~al.}{2013}]{switzer13}
{Switzer} E.~R.,  et~al., 2013, \mn@doi [\mnras] {10.1093/mnrasl/slt074}, \href
  {http://adsabs.harvard.edu/abs/2013MNRAS.434L..46S} {434, L46}

\bibitem[\protect\citeauthoryear{{Tanidis} \& {Camera}}{{Tanidis} \&
  {Camera}}{2019}]{2019MNRAS.489.3385T}
{Tanidis} K.,  {Camera} S.,  2019, \mn@doi [\mnras] {10.1093/mnras/stz2366},
  \href {https://ui.adsabs.harvard.edu/abs/2019MNRAS.489.3385T} {489, 3385}

\bibitem[\protect\citeauthoryear{Verde \& Matarrese}{Verde \&
  Matarrese}{2009}]{Verde:2009hy}
Verde L.,  Matarrese S.,  2009, \mn@doi [Astrophys. J.]
  {10.1088/0004-637X/706/1/L91}, 706, L91

\bibitem[\protect\citeauthoryear{{Villaescusa-Navarro}, {Viel}, {Datta}  \&
  {Choudhury}}{{Villaescusa-Navarro} et~al.}{2014}]{villa14}
{Villaescusa-Navarro} F.,  {Viel} M.,  {Datta} K.~K.,   {Choudhury} T.~R.,
  2014, eprint arXiv:1405.6713, \href
  {http://adsabs.harvard.edu/abs/2014arXiv1405.6713V} {}

\bibitem[\protect\citeauthoryear{{Zafar}, {P{\'e}roux}, {Popping}, {Milliard},
  {Deharveng}  \& {Frank}}{{Zafar} et~al.}{2013}]{zafar2013}
{Zafar} T.,  {P{\'e}roux} C.,  {Popping} A.,  {Milliard} B.,  {Deharveng}
  J.-M.,   {Frank} S.,  2013, \mn@doi [\aap] {10.1051/0004-6361/201321154},
  \href {http://adsabs.harvard.edu/abs/2013A%26A...556A.141Z} {556, A141}

\bibitem[\protect\citeauthoryear{{Zwaan}, {Meyer}, {Staveley-Smith}  \&
  {Webster}}{{Zwaan} et~al.}{2005a}]{zwaan05}
{Zwaan} M.~A.,  {Meyer} M.~J.,  {Staveley-Smith} L.,   {Webster} R.~L.,  2005a,
  \mn@doi [\mnras] {10.1111/j.1745-3933.2005.00029.x}, \href
  {http://adsabs.harvard.edu/abs/2005MNRAS.359L..30Z} {359, L30}

\bibitem[\protect\citeauthoryear{{Zwaan}, {van der Hulst}, {Briggs},
  {Verheijen}  \& {Ryan-Weber}}{{Zwaan} et~al.}{2005b}]{zwaan2005a}
{Zwaan} M.~A.,  {van der Hulst} J.~M.,  {Briggs} F.~H.,  {Verheijen} M.~A.~W.,
   {Ryan-Weber} E.~V.,  2005b, \mn@doi [\mnras]
  {10.1111/j.1365-2966.2005.09698.x}, \href
  {http://adsabs.harvard.edu/abs/2005MNRAS.364.1467Z} {364, 1467}

\makeatother
\end{thebibliography}



\appendix

\section{Useful formulae}
From the definition of the growth rate $f(z)$ and its parameterisation in terms of $\gamma$, we can express the growth factor according to
\begin{equation}
D(z)\propto\exp\left\{-\int_0^z\ud\tilde z\,\frac{\left[\Omega_{\rm m}(\tilde z)\right]^\gamma}{1+\tilde z}\right\},
\end{equation}
thus having
\begin{equation}
\frac{\partial\ln D}{\partial\gamma}=-\int_0^z\ud\tilde z\,\frac{\left[\Omega_{\rm m}(\tilde z)\right]^\gamma\ln\left[\Omega_{\rm m}(\tilde z)\right]}{1+\tilde z}.
\end{equation}

In our specific case, the analytical expressions used in \autoref{sec:MG} for the relevant quantities are
\begin{equation}
\nu f(\nu) = 2 A \left(1 + \frac{1}{\nu'^{2q}} \right)\left( \frac{\nu'^{2}}{2 \pi} \right)^{1/2} \exp(-\nu'^2/2),\label{eq:nufnu1}
\end{equation}
where $\nu' = \sqrt{a} \nu$, and the best fit parameters are $A = 0.3222$, $a = 0.707$ and $q = 0.3$ \citep{Sheth:2001dp}.
From this, $n_{\rm h}(M,z)$ is defined through
\begin{equation}
\nu f(\nu) = M^2 \frac{n_{\rm h}
(M,z)}{\bar{\rho}} \frac{\ud \ln M}{\ud \ln \nu},\label{eq:nufnu2}
\end{equation}
where $\bar{\rho}$ is the background density. By equating \autoref{eq:nufnu1} and \autoref{eq:nufnu2}, we can express \begin{equation}
n_{\rm h}(M,z)=2 A \left(1 + \frac{1}{\nu'^{2q}} \right)\left( \frac{\nu'^{2}}{2 \pi} \right)^{1/2} \exp(-\nu'^2/2)\frac{\bar{\rho}}{M}\frac{\ud\ln\sigma^{-1}}{\ud M},
\end{equation}
in which we made use of the definition of $\nu$.

Similarly, the expression for the halo bias is given by \citep{scoccimarro2001}
\begin{equation}
b_{\rm h} = 1 + \frac{\nu'^2 -1}{\delta_c(z)} + \frac{2q/\delta_c(z)}{1 + \nu'^{2q}} .
\end{equation}
where $\delta_c(z) \equiv \delta_c/D(z)$.

Using these forms, the expressions for the logarithmic derivatives become
\begin{equation}
\frac{\partial \ln n_h}{\partial \ln \nu} = -\frac{2q}{1 + (a \nu^2)^q} - a \nu^2 +1
\end{equation}
and
\begin{equation}
\frac{\partial \ln b_h}{\partial \ln \nu} =
2 \left\{
a \nu^2 - \frac{2 q^2 (a \nu^2)^q}{[1+(a \nu^2)^q]^2}
\right\}\left[\delta_c(z) - \frac{\partial \ln n_h}{\partial \ln \nu}\right]^{-1}.
\end{equation}

\section{Performance tests on the Fisher matrix procedure}
As described in the main text, the we follow the Fisher matrix procedure of \citet{Padmanabhan2017}, to which we add the beyond-\lcdm\ parameters of modified gravity and primordial non-Gaussianity. The code employed in the present analysis is the Fisher matrix module of \texttt{CCCP},\footnote{Camera's Code for Cosmological Perturbations \citep[see][Section~4.1]{2019arXiv191009273E}.} which has recently been validated within the internal, code-comparison effort of the \textit{Euclid} Consortium's Inter Science-working-group Task-force for Forecasting (IST:F). We find remarkable sub-percent level agreement between the results obtained with \texttt{CCCP} and the forecast constraints published in \citet{Padmanabhan2017}, where the stability of the derivatives with respect to astrophysical and (standard) cosmological parameters was extensively studied. The extensions considered here have likewise been variously tested and confirmed, in particular due to the fact that the derivatives with respect to both $\fnl$ and $\gamma$ are analytical.

A remark is worth being made on the only significant difference between the analysis carried out in this paper and the approach of \citet{Padmanabhan2017}. In that previous study, the Fisher matrices were constructed by summing over all redshift bins the elements of \autoref{eq:Fisher}, i.e. without binning in multipole space. In other words, the cumulative Fisher matrix, $\mathbfss F$, was built as in \autoref{eq:F_tot} but with the substitution $\Delta\ell_m\mathbfss F_{\bar{\ell}_m}(z_i)\to\mathbfss F_{\ell_m}(z_i)$. This was a conservative choice, made to avoid over-counting information from $\ell$-modes too close to each other, for which the Gaussian covariance estimate of \autoref{eq:Delta_Cl} might have been too simplistic an assumption. Nevertheless, we decide here to follow the more standard approach of modelling the multipole binning through the inclusion of the $N_\ell$ bin widths, $\Delta\ell_m$, whose log-centres are the $\bar{\ell}_m$'s. We justify our choice with the following considerations:

\begin{itemize}
\item First, being the current standard in the literature, it is thus more straightforward to compare our results to other work on the topic. For instance, our forecast marginal errors on cosmological parameters are in good agreement with the findings of \citet{Bull:2014rha}, despite the different approach followed in that paper.
\item Secondly, we have checked that the logarithmic binning we adopt does not introduce spurious information. In particular, we find that both logarithmic and linear binning, as well as summing all the multipoles from $\ell_{\rm min}$ to $\ell_{\rm max}$, does not make much difference to the forecasts. Noticeably, this holds true for all the cosmological and astrophysical parameters \textit{except} $\fnl$. In the case of $\fnl$, the marginal errors on this parameter are far worse using linear binning compared to logarithmic binning, and best when using all the multipoles. This is consistent with expectations, since most of the information on $\fnl$ is on ultra-large scales, so if one samples multipoles linearly, a lot of the information is lost, whilst one gains the most by including all multipoles.
\end{itemize}

\section{Results for M{\scriptsize eer}KAT}\label{sec:extraexp}

\begin{figure*}
\centering
\includegraphics[width=\textwidth]{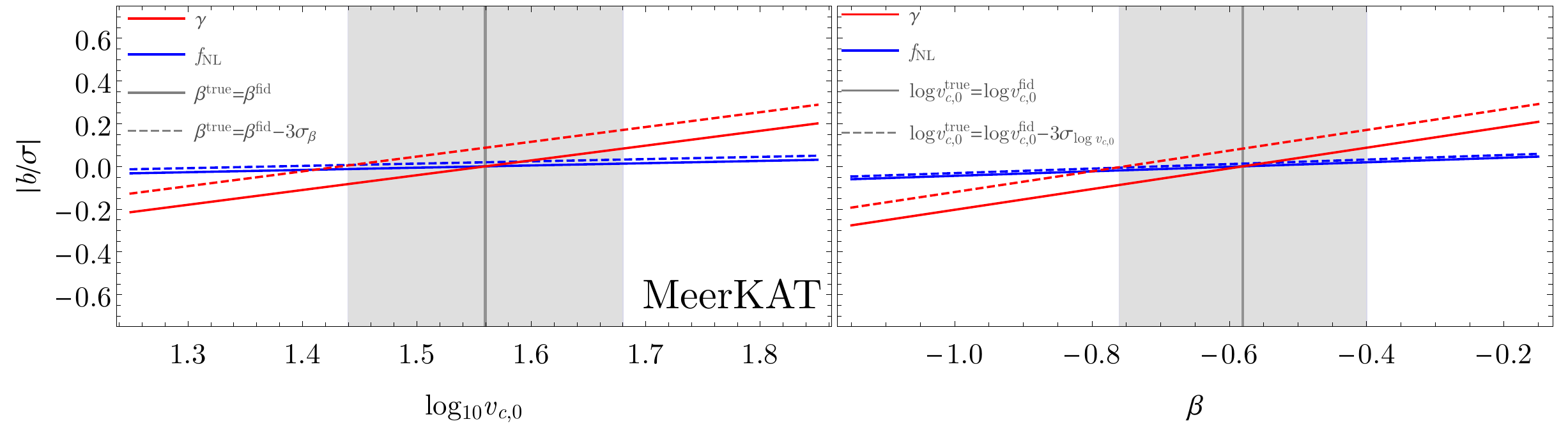}
\caption{Same as \autoref{fig:relbias_vc0} (left panel) and \autoref{fig:relbias_beta} (right panel), but for the case of MeerKAT.}
\label{fig:Fig12-CHIME}
\end{figure*}

\begin{figure*}
\centering
\includegraphics[width=\textwidth]{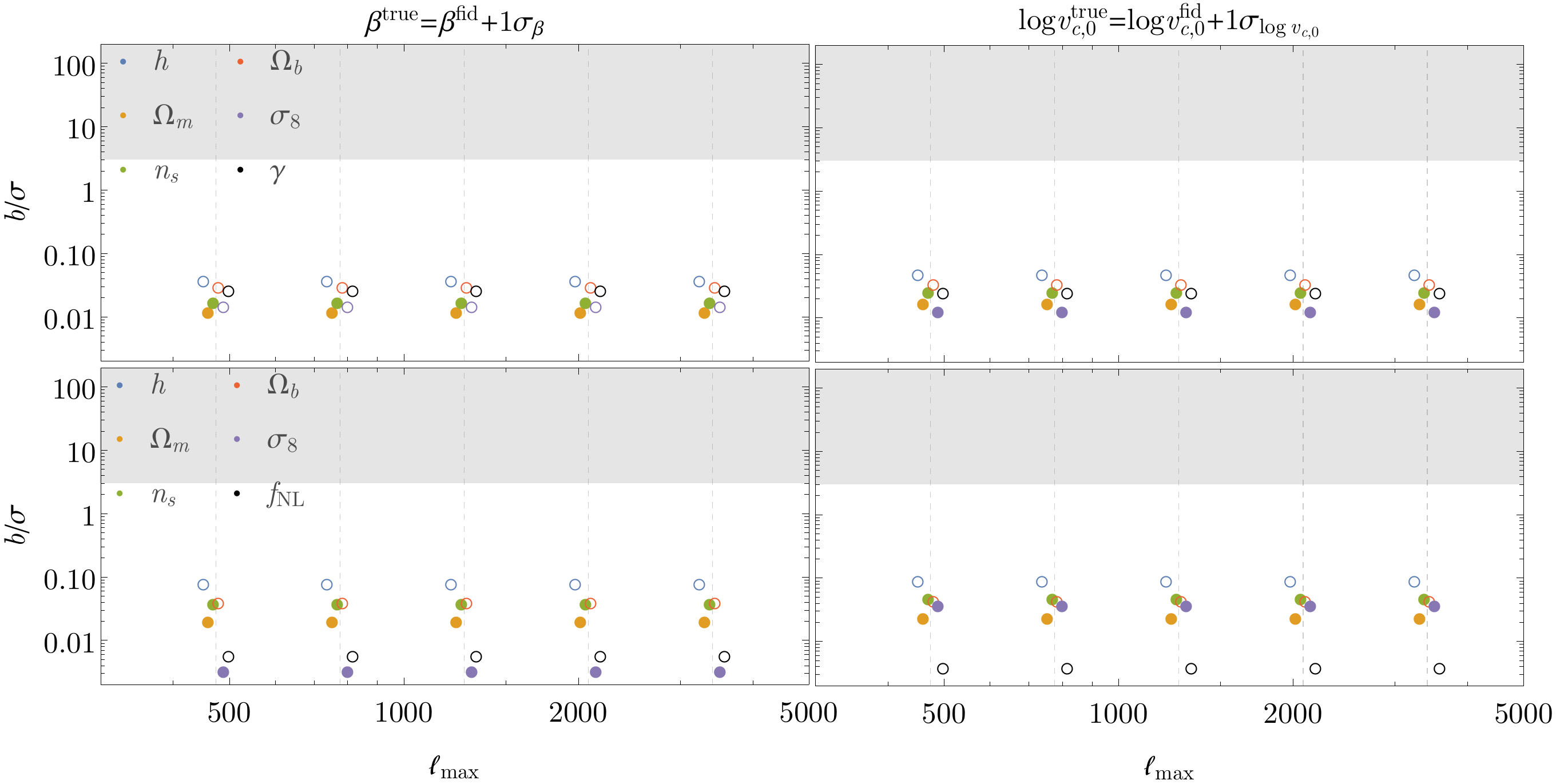}
\caption{Same as \autoref{fig:relbias3}, but for the case of MeerKAT.}
\label{fig:relbias2}
\end{figure*}

In the main text, we investigated the relative biases in recovery of the cosmological parameters including extensions to the standard \lcdm\ framework, using the fiducial configurations of CHIME and SKA1-MID to calculate the noise and redshift extents of the experiments. Here, we plot the same results as in Figures \ref{fig:relbias_vc0}, \ref{fig:relbias_beta} and \ref{fig:relbias3} of the main text, but now for the case of the MeerKAT-like experimental configuration. The results (shown in Figures \ref{fig:Fig12-CHIME} and \ref{fig:relbias2}) are qualitatively the same as those in SKA I MID, with overall differences in the relative bias due to the different constraining power when compared to SKA1-MID, as well as the different redshift range covered.

\bsp 
\label{lastpage}
\end{document}